\documentclass[a4paper,11pt]{article}
\usepackage{jheppub}

\pdfoutput=1
\RequirePackage{snapshot} % required by bundledoc
\usepackage{hyperref}
\usepackage{graphicx}
\usepackage[utf8]{inputenc}
\usepackage{xspace}
\usepackage{amsmath,amssymb}
\usepackage{bbm}
\usepackage{xcolor}
\usepackage{booktabs}
\usepackage{subcaption}
\usepackage{mathtools}

\newcommand{\order}[1]{{\cal O}\left(#1\right)}

\newcommand{\abar}{\bar{\alpha}}

\title{Next-to-leading non-global logarithms in QCD}

\author[a]{Andrea Banfi,}
\author[b]{Fr\'ed\'eric A. Dreyer,}
\author[c]{Pier Francesco Monni.}

\affiliation[a]{Department of Physics and Astronomy, University of
  Sussex, Sussex House, Brighton, BN1 9RH, UK}
\affiliation[b]{Rudolf Peierls Centre for Theoretical Physics,
  Clarendon Laboratory, Parks Road, Oxford OX1 3PU, UK}
\affiliation[c]{CERN, Theoretical Physics Department, CH-1211 Geneva 23, Switzerland}

\emailAdd{a.banfi@sussex.ac.uk}
\emailAdd{frederic.dreyer@physics.ox.ac.uk}
\emailAdd{pier.monni@cern.ch}

\preprint{OUTP-21-08P, CERN-TH-2021-040}

\abstract{Non-global logarithms arise from the sensitivity of collider
  observables to soft radiation in limited angular regions of phase
  space.
  Their resummation to next-to-leading logarithmic (NLL) order has
  been a long standing problem and its solution is relevant in the
  context of precision all-order calculations in a wide variety of
  collider processes and observables.
  In this article, we consider observables sensitive only to soft
  radiation, characterised by the absence of Sudakov double
  logarithms, and we derive a set of integro-differential equations
  that describes the resummation of NLL soft corrections in the
  planar, large-$N_c$ limit.
  The resulting set of evolution equations is derived in
  dimensional regularisation and we additionally provide a formulation
  that is manifestly finite in four space-time dimensions.
  The latter is suitable for a numerical integration and can be
  generalised to treat other infrared-safe observables sensitive
  solely to soft wide-angle radiation.
  We use the developed formalism to carry out a fixed-order
  calculation to ${\cal O}(\alpha_s^2)$ in full colour for both the
  transverse energy and energy distribution in the interjet region
  between two cone jets in $e^+e^-$ collisions. We find that the
  expansion of the resummed cross section correctly reproduces the
  logarithmic structure of the full QCD result.}

\keywords{}
\begin{document}
\setlength{\parskip}{0pt}
%\raggedbottom
\maketitle
\flushbottom

\section{Introduction}  
The accurate theoretical description of non-global QCD
observables~\cite{Dasgupta:2001sh,Dasgupta:2002bw,Banfi:2002hw} is among
the main obstacles on the path towards precision collider
phenomenology.
Non-global observables are commonly characterised by kinematic
constraints on limited angular regions of the radiation phase space,
and occur ubiquitously at colliders, for instance via the use of jets
or often when specific fiducial cuts are applied on
experimental measurements.
Such observables are sensitive to the coherent pattern of soft
radiation outside the measured region of phase space, and their
description therefore requires the calculation of such effects at all
perturbative orders in the strong coupling.
To achieve this resummation one meets two types of theoretical
challenges. 
Firstly, the presence of angular cuts in the definition of the
observable makes it impossible to handle the geometry of the problem
analytically. This is because the distribution of the soft radiation
in the angular region where the observable is defined (e.g.\ a jet cone
or a rapidity interval) depends on the full angular pattern of the
radiation in the event after the evolution from the hard scattering
scale down to the low scales at which the measurement is performed.
Secondly, the evolution itself is complicated because the colour
structure of the squared amplitude grows drastically with each new
soft emission.
The most striking consequence of such a peculiar structure is that
already at leading logarithmic (LL) accuracy, the standard
exponentiation of soft LL corrections does not hold, and one has to
solve a non-linear evolution equation, the Banfi-Marchesini-Smye (BMS)
equation~\cite{Dasgupta:2001sh,Dasgupta:2002bw,Banfi:2002hw}, to carry
out the resummation of logarithmically enhanced corrections.

Besides the relevance of non-global resummations for collider
phenomenology, a theoretical understanding of their dynamics is
instrumental in the context of developing more accurate parton-shower
algorithms (see e.g.\
Refs.~\cite{Dasgupta:2018nvj,Bewick:2019rbu,Dasgupta:2020fwr,Forshaw:2020wrq,Platzer:2020lbr,Hamilton:2020rcu,Nagy:2020rmk,Nagy:2020dvz,Karlberg:2021kwr,Dulat:2018vuy}
for recent work). Specifically, the resummation of next-to-leading
logarithmic (NLL) non-global logarithms is a crucial ingredient for
the development of NNLL algorithms, that are necessary to achieve
sufficiently accurate event simulation both at present and future
colliders.
Moreover, their study is also motivated by purely theoretical
interests, related to the discovery of a connection between the
evolution of non-global dynamics and the Balitsky-Kovchegov (BK)
equation~\cite{Weigert:2003mm,Hatta:2008st,Caron-Huot:2015bja}.

The resummation of LL non-global corrections in the large-$N_c$ limit
was formulated about 20 years ago in the seminal work of
Refs.~\cite{Dasgupta:2001sh,Dasgupta:2002bw,Banfi:2002hw}, and has
seen substantial interest in recent
years~\cite{Forshaw:2009fz,DuranDelgado:2011tp,Schwartz:2014wha,Becher:2015hka,Larkoski:2015zka,Becher:2016mmh,Becher:2016omr,Neill:2016stq,Caron-Huot:2016tzz,Larkoski:2016zzc,Becher:2017nof,Martinez:2018ffw,Balsiger:2018ezi,Neill:2018yet,Balsiger:2019tne,Balsiger:2020ogy}. Furthermore,
the authors of
Refs.~\cite{Hatta:2013iba,Hagiwara:2015bia,Hatta:2020wre} have
extended the LL resummation to include finite-$N_c$ corrections,
finding that subleading-colour corrections are numerically small in
common applications. Their study is however of paramount importance
for the understanding of the structure of super-leading logarithmic
corrections in non-global observables at hadron
colliders~\cite{Forshaw:2006fk,Forshaw:2008cq}.
The calculation of NLL corrections has
inspired a considerable amount of remarkable theoretical work along
the years, and formulations of the NLL resummation have been achieved
in different theoretical
formalisms~\cite{Becher:2015hka,Becher:2016mmh,Caron-Huot:2015bja}. However,
a full resummation of NLL corrections for a physical observable has
not yet been achieved.

In this article, we develop a formalism to resum non-global logarithms
at NLL accuracy in the large-$N_c$ limit. The resummation relies on a
set of non-linear evolution equations that can be solved numerically,
for instance by means of Monte Carlo (MC) simulations. 
We apply the developed formalism to the fixed order calculation of the
energy and transverse energy distribution in the region between two
cone jets in the process $e^+e^-\to$ 2 jets, and compare our findings
to an exact fixed-order prediction.
The evolution equations are derived in dimensional regularisation, and
later recast in a form that is manifestly finite in four dimensions,
hence making it very suitable for a numerical integration. 
The numerical solution of the proposed equations, as well as the corresponding
resummation of NLL corrections, will be presented in a forthcoming
publication.
The paper is structured as follows: Section~\ref{sec:formalism}
introduces the formalism used throughout the paper, and
Section~\ref{eq:RGEs} presents the strategy used to derive the
evolution equations. The detailed derivation of the NLL evolution
equation is discussed in Section~\ref{sec:NLL}, while
Section~\ref{sec:3jet} presents the calculation of the one-loop hard
matching coefficients necessary for a NLL calculation of the physical
cross section for the observables considered here. Finally, in
Section~\ref{eq:event2} we perform a fixed-order expansion up to
${\cal O}(\alpha_s^2)$ and compare our findings to the exact
calculation obtained with the program {\sc
  Event2}~\cite{Catani:1996vz}. Section~\ref{eq:conclusions} contains
our concluding remarks and outlook.

\section{Formalism and notation}
\label{sec:formalism}
\begin{figure}[htbp]
  \centering
  \includegraphics{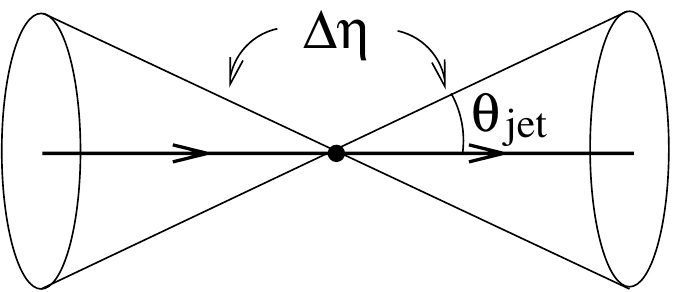}
  \caption{The rapidity slice where the measurement is performed.}
  \label{fig:obs}
\end{figure}

Let us consider the production of two jets in $e^+e^-$ annihilation at
a centre-of-mass energy $\sqrt{s}$. Considering the thrust axis as a
  reference axis, we define two jets in the two opposite hemispheres
  of the event by considering two cones of opening angle
  $\theta_{\rm jet}$ around the thrust axis. We focus on the rapidity
  region between the two cone jets, of total width
\begin{equation}
\Delta\eta \coloneqq \ln\frac{1+c}{1-c} \,,\qquad c=\cos \theta_{\rm jet}\,.
\end{equation}
We will informally refer to this region as the {\it rapidity slice}
centred at $\eta=0$ with respect to the thrust axis (see Fig.~\ref{fig:obs}).
In this paper, we study the distributions of both the energy ($E$) and
the transverse energy ($E_t$) in such a slice, and denote by
$\Sigma(v)$ the cumulative distribution for either observable to be
less than $v$, defined as follows
\begin{equation}
\Sigma(v)\coloneqq\frac{1}{\sigma_0}\int_{0}^v \frac{d\sigma}{d v^\prime} d v^\prime\,,
\end{equation}
where $\sigma_0$ is the Born cross section for $e^+e^-\to$ hadrons. In
the limit in which $v=\{E,E_t\}$ is small, large logarithms
$L=\ln(\sqrt s/v)$ spoil the convergence of fixed-order perturbative
expansions and must be resummed at all perturbative orders.
Since this observable is affected only by soft emissions at wide
angles, the largest logarithms in $\Sigma(v)$, the leading logarithms,
are of the form $\alpha_s^n L^n$. For $\alpha_s L\sim 1$, all terms
suppressed by an extra power of $\alpha_s$ give next-to-leading
logarithmic (NLL, $\alpha_s^n L^{n-1}$) contributions, and so on.
The phase-space constraint for the observables we consider admits a factorised
expression in Laplace space of the type
\begin{equation}
\label{eq:observable}
\Theta\left(v - \sum_{2|\eta_i| \leq \Delta\eta} v(k_i)\right) = \frac{1}{2\pi i}
\int_{\gamma}\frac{d\nu}{\nu} e^{\nu v} \prod_{2|\eta_i| \leq \Delta\eta} e^{-\nu v(k_i)} \coloneqq \frac{1}{2\pi i}
\int_{\gamma}\frac{d\nu}{\nu} e^{\nu v} \prod_i u(k_i)\,,
\end{equation}
where $v(k_i) = \{\omega_i , |\vec{k}_{ti}|\}$ for $v=\{E,E_t\}$
respectively. Here $\vec{k}_{ti}$ is the transverse momentum of
particle $k_i$ with respect to the thrust axis, $\omega_i$ is its
energy in the lab frame, and we defined $u(k_i)$ as the {\it source}
corresponding to the measurement. It takes the form
\begin{equation}
\label{eq:source}
u(k) = \Theta_{\rm out}(k) + \Theta_{\rm in}(k) e^{-\nu v(k) } \,,
\end{equation}
where the trigger function $\Theta_{\rm in}(k)$
($\Theta_{\rm out}(k)$) is 1 if the particle $k$ is inside (outside) a
rapidity slice of total width $\Delta\eta$, and zero otherwise. The
contour $\gamma$ lies parallel to the imaginary axis to the right of
all singularities of the integrand.

Without any emissions, at the lowest order in perturbation theory,
$\Sigma(v)=1$, and the event is made up of a quark of momentum $p_1$
and an antiquark of momentum $p_2$, back-to-back and aligned along the
thrust axis. When extra radiation is considered, $\Sigma(v)$ can be
expressed as
\begin{equation}
\label{eq:master}
\Sigma(v) \coloneqq \sum_{n=2}^{\infty}{\cal H}_n\otimes S_n(v) = {\cal H}_2\otimes S_2(v) + {\cal H}_3\otimes S_3(v) +\cdots
\end{equation}
where the hard factors 
\begin{equation}
{\cal H}_n\coloneqq {\cal H}_{1\dots n}
\end{equation}
describe configurations with $n$ hard QCD partons along the light-like
directions $n_1,\dots,n_n$ (with $n^2_i=0$ and $|\vec{n}|^2=1$), while
the soft factors 
\begin{equation}
S_n \coloneqq S_{1\dots n}
\end{equation}
describe the emission of soft radiation off a hard system with $n$
hard emitters along the same directions. The convolutions in
Eq.~\eqref{eq:master} are meant to indicate that the directions of the
hard emitters in the hard and soft factors are the same, namely
\begin{equation}
\label{eq:convolution}
{\cal H}_n\otimes S_n(v) = \int\left(\prod_{i=i}^n d^{2}\Omega_i\right)
{\cal H}_{1\dots n} \times S_{1\dots n} (v),
\end{equation}
where $\Omega_i$ indicates the solid angle of the $i$-th hard emitter,
namely the direction of the $\vec{n}_i$ vector, specified by a
longitudinal ($\theta$) and an azimuthal ($\phi$) angle.
Each of the above ingredients admits a perturbative expansion in the
strong coupling constant
\begin{align}
\label{eq:conventions}
 {\cal H}_n = \sum_{i=n-2}^\infty\frac{\alpha_s^i}{(2\pi)^i}  {\cal
  H}_n^{(i)},\qquad
 S_n = \sum_{i=0}^\infty\frac{\alpha_s^i}{(2\pi)^i} S_n^{(i)},
\end{align}
where
${\cal H}_2^{(0)}=\delta(\cos\theta_1-1) \delta(\cos\theta_2+1)
\delta(\phi_1) \delta(\phi_2)$ and $S_n^{(0)}=1$.
At LL accuracy, the resummation for each observable requires only the
first term in the r.h.s.\ of Eq.~\eqref{eq:master}, with the leading
order ${\cal H}_2$ and the LL soft factor $S_2$ whose evolution is
governed by the BMS equation~\cite{Banfi:2002hw}.
At NLL one needs to include both the first
($ {\cal H}_2\otimes S_2(v)$) and second
($ {\cal H}_3\otimes S_3(v) $) term on the r.h.s.\ of
Eq.~\eqref{eq:master}, where $ {\cal H}_2$ is to be computed at
next-to-leading order and $ {\cal H}_3$ at leading order.
Each of the latter hard factors is individually infrared finite. The
large logarithms of $v$ are entirely contained in the soft factors
$S_n$, and all convolutions defined in Eq.~\eqref{eq:convolution} can
now be carried out in $2$ dimensions.
At this point we can achieve NLL accuracy by including the
${\cal O}(\alpha_s)$ corrections to ${\cal H}_2$ and ${\cal H}_3$, as
well as the soft factor $S_3$ at LL and the soft factor $S_2$ at NLL.
In the following sections we will define each of the above
contributions in detail.

\section{Evolution equations for the soft factors in dimensional
  regularisation}
\label{eq:RGEs}
We start by deriving the evolution equation for the soft factors $S_n$
that appear in Eq.~\eqref{eq:master}.
Given the complexity of the problem due to the fast growth of the
colour space, in the following we will work in the widely used
large-$N_c$ limit, where all real and virtual graphs are considered in
the planar limit. Beyond this limit, the LL calculation has been
performed only in a few selected
cases~\cite{Hatta:2013iba,Hagiwara:2015bia}. The crucial advantage of
using the large-$N_c$ limit is that it makes it possible to write
closed evolution equations in terms of colour dipoles. This in turn
makes them suitable for Monte-Carlo integration.
We will first re-derive the LL evolution equation formulated in
Ref.~\cite{Banfi:2002hw}, and we then move on to derive the new NLL
evolution equations that constitutes one of the main results of this
paper.

\subsection{The LL (BMS) evolution equation for $S_2$ and $S_3$}
To derive the evolution equations, it is convenient to work with the
Laplace transform of the soft factors of Eq.~\eqref{eq:master}, as the
observable takes a factorised form in this space. We thus define the
Laplace-space soft factors $G_{1 2\cdots n}$ as
\begin{equation}
\label{eq:laplace}
  S_{n}(v) = \int_{\gamma}\frac{d\nu}{2\pi i \nu} e^{\nu v} G_{12\cdots
    n}[Q;u]\,.
\end{equation}
We then start from an initial state made of the quark-antiquark pair
$\{p_1,p_2\}$, which in a large-$N_c$ picture defines the radiating
colour dipole $\{12\}$. When the emission of extra soft gluons
strongly ordered in energy is considered, the initial dipole receives
radiative corrections according to the squared amplitude (see
e.g.~\cite{Bassetto:1984ik,Fiorani:1988by})
\begin{equation}
\label{eq:SO}
{\cal A}_{12}^2 = \abar^n(\mu)(2\pi)^{2 n}(\mu^{2\epsilon})^
  n\sum_{\pi_n}\frac{(p_1 \cdot p_2)}{(p_1 \cdot k_{i_1})(k_{i_1} \cdot k_{i_2})\dots(k_{i_n} \cdot p_2)}\,,
\end{equation}
where the sum runs over all $n!$ permutations, and we defined
$\abar= N_c \alpha_s(\mu)/\pi$, where $\alpha_s$ is the QCD coupling
in the $\overline{\rm MS}$ scheme. This approximation is valid at the
leading logarithmic order, and we will consider higher-order
corrections in the next section.
We will derive the evolution equation using a branching
formalism. This is based on identifying a resolution variable $Q$,
such that subsequent evolution steps in $Q$ reproduce the correct
squared amplitude at a given logarithmic accuracy. Moreover, it is of
course necessary that at fixed $Q$, the contribution of the soft
evolution to the cross section be infrared and collinear (IRC)
finite. The choice of the resolution scale $Q$ can vary for different
problems (it can for instance coincide with the
virtuality~\cite{Catani:1992ua} of the radiating system or with the
relative angle~\cite{Dasgupta:2014yra} of the radiation w.r.t.\ the
emitter). For the problem at hand, the LL corrections originate from
radiation strongly ordered in energy. It is therefore natural to chose
$Q$ as the energy of the {\it hardest} gluon~\cite{Banfi:2002hw} and
introduce the real-emission contribution to $G_{12}[Q;u]$ defined by
\begin{equation}
\label{eq:LLreal}
G^{(\mathrm{R})}_{12}[Q;u] = \sum_n \frac{1}{n!}\int \left(\prod_i [d k_i]  u(k_i)\Theta(Q - \omega_i)\right){\cal
  A}_{12}^2 \,,
\end{equation}
where $\omega_i$ is the energy of gluon $k_i$. The evolution in $Q$
from low scales up to $Q\sim \sqrt{s}$ will achieve the resummation of
LL corrections. 
We also introduce the phase-space measure in $d=4-2\epsilon$ dimensions
\begin{equation}
[d k_i] \coloneqq \omega^{1-2\epsilon}_i\, d \omega_i\,
\frac{d^{2-2\epsilon}\Omega_i}{2 (2\pi)^{3-2\epsilon}}\,,
\end{equation}  
where $\Omega_i$ denotes the solid angle.
The inclusion of virtual corrections will be discussed shortly. 
The definition of $Q$ we just adopted is only collinear safe in the
strongly-ordered energy limit relevant for LL. However, our aim is to
formulate a NLL evolution equation that is well defined for any value
of $Q$ and for different choices of the IRC safe observable's source
$u$.
This requires modifying the definition of $Q$ in configurations with
two gluons with commensurate energy becoming collinear, crucial for
attaining NLL accuracy. We will come back to this point in
Section~\ref{sec:NLL}.

To predict how the multi-gluon system evolves with the scale $Q$, we
derive an evolution equation that describes the dependence of the soft
system on the radiation's energy. We start by considering the
large-$N_c$ real-emission contribution $G^{(\mathrm{R})}_{12}[Q;u]$
defined above.
We introduce the tree-level eikonal kernel
\begin{equation}
w^{(0)}_{12}(k) = 8\pi^2\frac{\mu^{2\epsilon}}{k_t^2} \,,\quad k_t^2 =
2\frac{(p_1\cdot k)(k\cdot p_2)}{(p_1\cdot p_2)}
\,,
\end{equation}
where $k_t$ is the transverse momentum with respect to the emitting
dipole. In the following we will work in the $\overline{\rm MS}$
scheme, defined from the bare coupling as
\begin{equation}
\label{eq:MS}
\alpha_s\mu^{2\epsilon}\to
\alpha_s(\mu)\mu^{2\epsilon}\frac{e^{\gamma_E\epsilon}}{(4\pi)^\epsilon}\left(1-\frac{\beta_0}{\epsilon}\alpha_s(\mu)+\dots\right)\,,
\end{equation}
where $\beta_0$ is the first coefficient of the QCD $\beta$ function
\begin{equation}
  \label{eq:beta0_largeN}
  \beta_0=\frac{11 C_A-2 n_f}{12\pi} \xrightarrow[]{N_c\gg 1} \frac{11
  }{12}\frac{N_c}{\pi}\coloneqq \bar{\beta}_0 \frac{N_c}{\pi}\,.
\end{equation}
In the large-$N_c$ limit, the factorisation properties of the squared
amplitude~\eqref{eq:SO} lead to the following evolution equation for
the real contribution
\begin{equation}
\label{eq:G0R-evolution}
Q \partial_Q G^{(\mathrm{R})}_{12}[Q;u] = \int [d k_a] \abar(k_{ta}) \,w^{(0)}_{12}(k_a) G^{(\mathrm{R})}_{1a}[Q;u]
  G^{(\mathrm{R})}_{a 2}[Q;u] u(k_a) Q \delta(Q-\omega_a)\,,
\end{equation}
where $ G^{(\mathrm{R})}_{1a}$ and $G^{(\mathrm{R})}_{a2}$ are defined
according to Eq.~\eqref{eq:SO} by just replacing either $p_1$ or $p_2$
with $k_a$. 
In this form, Eq.~\eqref{eq:G0R-evolution} is manifestly collinear
unsafe, and one needs to introduce the appropriate virtual
corrections. This can be done by imposing unitarity, which enforces
$G_{ij}[u]=1$ for $u(k_a)=1$ and for any value of $i$ and $j$ (in this
case $i,j=1,2,a$). We then obtain the final LL evolution equation for
the physical $G_{12}$ distribution, which reads
\begin{equation}
\label{eq:LL-evolution}
Q \partial_Q G_{12}[Q;u] = \int [d k_a] \abar(k_{ta}) \,w^{(0)}_{12}(k_a)\left(G_{1a}[Q;u]
  G_{a2}[Q;u] u(k_a) - G_{12}[Q;u]\right) Q \delta(Q-\omega_a)\,.
\end{equation}
The second term in the r.h.s.\ of the above evolution equation encodes
the LL contribution of a virtual gluon $k_a$.
The above equation is the BMS equation. With the boundary condition
\begin{equation}
\label{eq:initial-cond}
G_{12}[Q;u] = 1~{\rm for}~ Q = 0\,,
\end{equation}
and the normalisation $G_{12}[Q;1]=1$, this equation resums the
logarithmic terms $\ln(Q\nu)$ at LL accuracy.

Eq.~\eqref{eq:initial-cond} is well defined in dimensional
regularisation as the Landau singularity can be avoided by
analytically continuing $G_{12}$ to complex $\epsilon$
values~\cite{Magnea:2000ss} (albeit with $\Re(\epsilon) < 0$). When
taking the four-dimensional limit of Eq.~\eqref{eq:LL-evolution}, some
extra considerations are necessary and will be discussed in
Section~\ref{sec:4D}.
When the inverse Laplace transform~\eqref{eq:laplace} is considered,
this provides a resummation of the logarithms $L=\ln(Q/v)$ in
$\Sigma(v)$.

We can further manipulate Eq.~\eqref{eq:LL-evolution} and perform a
change of evolution variable from energy to the transverse momentum of
the soft radiation. At the leading (single) logarithmic level, we can
replace $\delta(Q-\omega_a)$ with $\delta(Q-k_{ta})$, where $k_{ta}$
is the transverse momentum of $k_a$ with respect to the emitting
dipole $\{12\}$. This is because for soft radiation emitted with a
large angle in the event centre-of-mass frame one has
$k_{ta}\sim\omega_a$ up to a ${\cal O}(1)$ function of the
pseudo-rapidity of the radiation. The latter function gives only rise
to NLL corrections, which are entirely accounted for by the source
$u(k)$. This gives
\begin{equation}
\label{eq:LL-evolution-kt-diff}
Q \partial_Q G_{12}[Q;u] = \int [d k_a] \abar(k_{ta}) \,w^{(0)}_{12}(k_a)\left(G_{1a}[Q;u]
  G_{a2}[Q;u] u(k_a) - G_{12}[Q;u]\right) Q \delta(Q-k_{ta})\,,
\end{equation}
with boundary condition given in Eq.~\eqref{eq:initial-cond}.

We now rewrite eq.~(\ref{eq:LL-evolution-kt-diff}) in a physically
appealing integral form. Let us first introduce the Sudakov form
factor $\Delta_{12}(Q)$
\begin{equation}
\label{eq:sudakov-LL}
\ln \Delta_{12}(Q) = -\int [dk] \abar(k_t) w_{12}^{(0)}(k)\Theta(Q-k_t)\,,
\end{equation}
where the phase space measure as a function of $k_t$ and the rapidity
$\eta$ with respect to the emitting dipole $\{12\}$ is given
by
\begin{equation}
\label{eq:phase-space-kt}
[d k]\coloneqq \frac{d \eta}{2}\frac{d^{2-2\epsilon}k_t}{(2\pi)^{3-2\epsilon}}\,.
\end{equation}
The upper bound of the rapidity integral can be consistently expanded
around its soft limit as
\begin{equation}
\label{eq:rapidity-bound}
|\eta| \leq \cosh^{-1}\left(\frac{\sqrt{2\, p_1\cdot p_2}}{2 k_t}\right)=\ln
\frac{\sqrt{2 \,p_1\cdot p_2}}{k_t}+{\cal O}\left(\frac{k_t^2}{2 \,p_1\cdot p_2}\right)\,,
\end{equation}
where we neglect ${\cal O}(k_t^2)$ terms as they only give rise to
subleading power corrections.
Eq.~\eqref{eq:LL-evolution-kt-diff} can then be written as
\begin{equation}
\label{eq:LL-evolution-kt-int}
Q \partial_Q \frac{G_{12}[Q;u]}{\Delta_{12}(Q)} = \int [d k_a] \abar(k_{ta}) \,\frac{w^{(0)}_{12}(k_a)}{\Delta_{12}(Q)}G_{1a}[k_{ta};u]
  G_{a2}[k_{ta};u] u(k_a) Q \delta(Q-k_{ta})\,.
\end{equation}
We can now bring the previous equation into an integral form as
\begin{equation}
\label{eq:LL-evolution-kt}
G_{12}[Q;u] =\Delta_{12}(Q)+ \int [d k_a] \abar(k_{ta}) \,w^{(0)}_{12}(k_a)\frac{\Delta_{12}(Q)}{\Delta_{12}(k_{ta})}G_{1a}[k_{ta};u]
  G_{a2}[k_{ta};u] u(k_a) \Theta(Q-k_{ta})\,,
\end{equation}
which can be solved iteratively. The infrared singularities in
Eq.~\eqref{eq:LL-evolution-kt} are regulated by dimensional
regularisation.
Imposing unitarity, i.e.\ setting all sources to 1 in
Eq.~\eqref{eq:LL-evolution-kt-int} gives
\begin{equation}
\label{eq:unitarityLL}
1 =\Delta_{12}(Q)+ \int [d k_a] \abar(k_{ta}) \,w^{(0)}_{12}(k_a)\frac{\Delta_{12}(Q)}{\Delta_{12}(k_{ta})} \Theta(Q-k_{ta})\,.
\end{equation}
The above equation identifies the Sudakov form factor $\Delta_{12}(Q)$
with the no-emission probability, and the second term on the
right-hand side represents the total probability for the emission of
one gluon. This physical interpretation of the Sudakov form factor
will be instrumental later when deriving the NLL evolution equation.
We note that the $k_t$-ordered formulation we have just introduced is
significantly different from the energy- ($\omega$-)ordered case of
Ref.~\cite{Banfi:2002hw} in the collinear limit, namely when $k_a$ is
emitted collinear to either of the dipole ends $p_1$ or $p_2$.
Specifically, the difference appears in a situation in which $k_a$ is
collinear to, say, $p_2$, and hence $k_{ta}\to 0$. This will
automatically also introduce an exponential suppression for all
emissions off the dipole $\{1a\}$ that still has a significant angular
phase space available for further emissions. This suppression is in
fact not present in the energy-ordered case, as long as $\omega_a$ is
different from zero.
In problems sensitive to soft radiation only, such as the one
considered in this article, this limit is entirely irrelevant. This is
because when $k_a$ is collinear to one of the dipole ends one has
$u=1$ (and hence $G=1$), giving no contribution to the observable.
This ensures that for the problem at hand working in terms of dipole
transverse momenta rather than energies is equivalent. However,
problems with sensitivity to collinear radiation (such as observables
with Sudakov double logarithms) would present some extra subtleties
and some care must be taken in this respect. As stressed multiple
times in this paper, we do not consider this class of observable here.

Before moving on with the NLL evolution, we discuss the evolution
properties of the soft factor $S_3$ defined on three light-cone
directions $n_1$, $n_2$, $n_3$. According to Eq.~\eqref{eq:laplace},
its Laplace transform is defined by
\begin{equation}
  S_{3}(v) = \int_{\gamma}\frac{d\nu}{2\pi i \nu} e^{\nu v} G_{123}[Q;u]\,.
\end{equation}
In the planar (large$-N_c$) limit, the emission of a hard gluon $p_3$
off the initial dipole defined by the quark and anti-quark momenta
$p_1$ and $p_2$ creates two adjacent dipoles $\{13\}$ and $\{32\}$. In
this limit, these two dipoles radiate incoherently, and therefore one
can write
\begin{equation}
G_{123} \xrightarrow[]{N_c\gg 1} G_{13}\, G_{32}\,.
\end{equation}
The above replacement is exact in the large$-N_c$ limit, and therefore
valid at all logarithmic orders. The evolution of each of the factors
in the r.h.s.\ of the above equation is described by
Eq.~\eqref{eq:LL-evolution-kt}. One final aspect that we need to
discuss is the initial scale of the evolution for $G_{13}$ and
$G_{32}$. In this case, the hard gluon $p_3$ carries a significant
fraction of the centre of mass energy $\sqrt{s}$. Therefore, in the
transverse momentum ordered picture considered here, any subsequent
soft emission will have a dipole transverse momentum smaller than that
of $p_3$ with respect to the $\{12\}$ dipole. We therefore set the
initial scale for the evolution of the $\{13\}$ and $\{32\}$ dipoles
to $p_{t3}$ where
\begin{equation}
p_{t3}^2=2\frac{(p_1\cdot p_3)(p_3\cdot  p_2)}{(p_1\cdot  p_2)}\,.
\end{equation}
We now observe that $p_{t3}\sim Q$, since $S_3$ is convoluted with
${\cal H}_{3}$ which vanishes by definition in the soft limit, that is
$p_{t3}\ll Q$. Therefore, up to subleading (NNLL) corrections we can
expand $p_{t3}$ about $Q$ and write
\begin{equation}
\label{eq:factS3}
G_{123}[Q;u] \xrightarrow[]{N_c\gg 1} G_{13}[Q;u]\, G_{32}[Q;u]\,.
\end{equation}

\subsection{The NLL evolution equation for $S_2$}
At the NLL order, the evolution
equation~\eqref{eq:LL-evolution-kt-diff} receives radiative
corrections to the kernel for the evolution of the soft factor
$S_2$. Conversely, as stressed in the previous section, the soft
factor $S_3$ obeys the factorisation~\eqref{eq:factS3} into two
independent colour dipoles, each of which evolves according to the LL
evolution equation~\eqref{eq:LL-evolution-kt-diff}.
Given the technical nature of the derivation of a NLL evolution
equation, we explain its structure in this section, and leave the
detailed derivation to Section~\ref{sec:NLL} for the interested
reader.
We start by considering the LL evolution
equation~\eqref{eq:LL-evolution-kt-diff} 
\begin{equation}
\label{eq:LL-evolution-kt-diff-symbolic}
Q \partial_Q G_{12}[Q;u] = {\mathbb K}^{\rm LL}[G[Q,u],u]\,,
\end{equation}
where we introduced a short hand notation for the LL evolution kernel
\begin{equation}
\label{eq:LLkernel}
{\mathbb K}^{\rm LL}[G[Q,u],u]\coloneqq\int [d k_a] \abar(Q) \,w^{(0)}_{12}(k_a)\left(G_{1a}[Q;u]
  G_{a2}[Q;u] u(k_a) - G_{12}[Q;u]\right) Q \delta(Q-k_{ta}).
\end{equation}
From the previous equation, we see that the LL evolution kernel is
made of a term which describes the real emission of a soft gluon, and
the corresponding virtual corrections at one loop. The combination of
the two is finite for an infrared safe observable described by the
source $u(k)$.  We note that derivative of $G_{12}$ with respect to
$\ln Q$ in Eq.~\eqref{eq:LL-evolution-kt-diff-symbolic} singles out,
by construction, the contribution of the {\it hardest} of the soft
gluons in real configurations. Conversely, virtual corrections are
included by unitarity as discussed above.
At NLL, we need to compute the radiative corrections to the r.h.s.\ of
Eq.~\eqref{eq:LL-evolution-kt-diff-symbolic}. 
At this order the $\ln Q$ derivative will resolve at most two
emissions with comparable transverse momentum (or equivalently,
energy) as well as the virtual corrections to the emission of the
hardest gluon considered in the LL kernel~\eqref{eq:LLkernel}.
Specifically, as shown in detail in Section~\ref{sec:NLL}, these
corrections can be all computed from known amplitudes. We can then
parametrise the NLL evolution equation as follows
\begin{align}
\label{eq:NLL-evolution-kt-diff-symbolic}
Q \partial_Q G_{12}[Q;u] &={\mathbb K}^{\rm NLL}[G[Q,u],u]\notag\\
&\coloneqq {\mathbb K}^{\rm RV+VV}[G[Q,u],u]+{\mathbb K}^{\rm RR}[G[Q,u],u]-{\mathbb K}^{\rm DC}[G[Q,u],u]\,.
\end{align}
In Eq.~\eqref{eq:NLL-evolution-kt-diff-symbolic},
${\mathbb K}^{\rm RV+VV}$ contains the purely virtual corrections to the
$\{12\}$ dipole up to two loops, as well as the real-virtual
corrections to the soft current up to one loop. Similarly,
${\mathbb K}^{\rm RR}$ contains the double real corrections describing
the emission of two unordered soft partons off the $\{12\}$ dipole.
Finally, the extra term ${\mathbb K}^{\rm DC}$ has the role of
subtracting the first iteration of the LL kernel~\eqref{eq:LLkernel},
so as to ensure a correct subtraction of the double counting at all
perturbative orders.

In the next section we will address the calculation of each of these
three ingredients necessary to formulate a complete NLL evolution
equation. The derivation will be carried out in $d=4-2\epsilon$
dimensions, but we will also present a simple procedure to perform a
local subtraction of the IRC divergences and formulate each of the
three contributions to the NLL
kernel~\eqref{eq:NLL-evolution-kt-diff-symbolic} in a form that is
manifestly finite in $d=4$ dimensions. The final results will be given
in Eqs.~\eqref{eq:Uvirtual},~\eqref{eq:Ureal},~\eqref{eq:Udc}.

\section{Derivation of the NLL evolution equation}
\label{sec:NLL}
We now extend the evolution equation~\eqref{eq:LL-evolution-kt} to
NLL accuracy. We stick to the transverse momentum with respect to the emitting
dipole as our evolution variable, although one could alternatively use
energy as done originally, which would lead to a slightly different
form of the final equation, with solutions identical up to subleading
logarithmic terms.

It is instructive to start by performing the first iteration of
Eq.~\eqref{eq:LL-evolution-kt}, and expand in a fixed number of
emissions. We obtain
\begin{align}
\label{eq:LL-evolution-kt-iterated}
G_{12}[Q;u] &=\Delta_{12}(Q)+ \int [d k_a] \abar(k_{ta}) \,w^{(0)}_{12}(k_a)\frac{\Delta_{12}(Q)}{\Delta_{12}(k_{ta})}\Delta_{1a}(k_{ta})
  \Delta_{a2}(k_{ta}) u(k_a) \Theta(Q-k_{ta})\notag\\
&+\int [d k_a] \int [d k_b] \abar(k_{ta}) \abar(k_{tb})
  \,w^{(0)}_{12}(k_a) w^{(0)}_{1a}(k_b)
\frac{\Delta_{12}(Q)}{\Delta_{12}(k_{ta})}\frac{\Delta_{1a}(k_{ta})}{\Delta_{1a}(k_{tb})}\Theta(k_{ta}-k_{tb})\notag\\
&\times
 \Delta_{1b}(k_{tb}) \Delta_{ba}(k_{tb})\Delta_{a2}(k_{ta}) u(k_a)u(k_b)\Theta(Q-k_{ta})\notag\\
&+\int [d k_a] \int [d k_b] \abar(k_{ta}) \abar(k_{tb})
  \,w^{(0)}_{12}(k_a) w^{(0)}_{a2}(k_b)
\frac{\Delta_{12}(Q)}{\Delta_{12}(k_{ta})}\frac{\Delta_{a2}(k_{ta})}{\Delta_{a2}(k_{tb})}\Theta(k_{ta}-k_{tb})\notag\\
&\times
 \Delta_{1a}(k_{ta}) \Delta_{ab}(k_{tb})\Delta_{b2}(k_{tb})
  u(k_a)u(k_b) \Theta(Q-k_{ta})+ {\rm \{\geq 3\,emissions\}}\,,
\end{align}
where we neglected terms describing the emission of more than two
gluons in large-$N_c$ limit. With a slight abuse of notation, we now denoted
with $k_{ta}$ the usual transverse momentum of $k_a$ with respect to
the $\{12\}$ dipole, and with $k_{tb}$ that of $k_b$ with respect to
the emitting dipole, either $\{1a\}$ or $\{a2\}$, in the last two
terms of Eq.~\eqref{eq:LL-evolution-kt-iterated}, respectively.
 Explicitly, each theta function has to be interpreted according to the following definition:
  \begin{equation}
    \label{eq:ktij-def}
    w^{(0)}_{ij}(k_b) \Theta(k_{ta}-k_{tb}) \coloneqq w^{(0)}_{ij}(k_b) \Theta(k_{ta}-k^{(ij)}_{tb})\,,
  \end{equation}
where $k^{(ij)}_{tb}$ is the transverse momentum of $k_b$ with respect to the ``emitting'' dipole $\{ij\}$.
The first line of Eq.~\eqref{eq:LL-evolution-kt-iterated} encodes the
single real emission at tree level, while the last four lines encode
the double real correction in the two colour flows that contribute to
the large-$N_c$ pattern. However, these corrections are only accounted
for in the strongly ordered limit in
Eq.~\eqref{eq:LL-evolution-kt-iterated}, and we must instead use the
full unordered limit to account for NLL corrections.

There are two types of contributions that arise when we consider the
corrections to the squared amplitude of Eq.~\eqref{eq:SO}. Eq.~\eqref{eq:SO}
is obtained in the limit of emissions strongly ordered in energy, or
equivalently in dipole transverse momentum (we recall that we work in
the limit of soft radiation emitted with wide angle w.r.t.\ the Born
legs). This limit leads to the leading terms $(\alpha_s L)^n$ resummed
by Eq.~\eqref{eq:LL-evolution-kt}. We start by noticing that in the
kernel of the LL evolution equation~\eqref{eq:LL-evolution-kt}, the
eikonal squared amplitude $w^{(0)}_{12}(k_a)$, describes the emission
of the soft gluon which carries the largest transverse momentum
$k_{ta}$ with respect to the emitting dipole.  To gain control over
NLL terms of order $\alpha_s(\alpha_s L)^n$, we need to account for
the corrections of relative order ${\cal O}(\alpha_s)$ to the above
kernel. These are discussed in the following.

\begin{figure}[htbp]
  \centering
  \includegraphics[width=.8\textwidth]{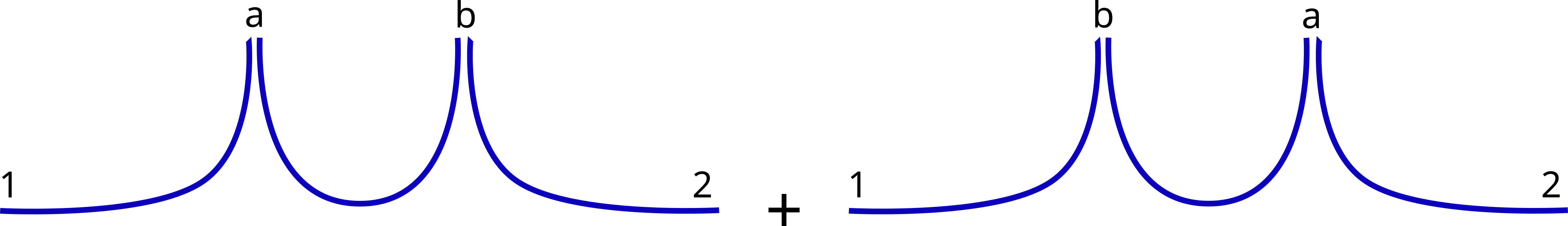}
  \caption{The large-$N_c$ colour-flow decomposition of the double real squared
    amplitude.}
  \label{fig:dip_flows}
\end{figure}

\paragraph{Real corrections.} The squared of the double soft current,
which is needed to describe the emission of two soft partons of
commensurate energy, can be split into two terms in the large-$N_c$
limit as
$\tilde{w}^{(0)}_{12}(k_a,k_b) + \tilde{w}^{(0)}_{12}(k_b,k_a)$, each
of which corresponds to different colour flows as shown schematically
in Fig.~\ref{fig:dip_flows}.

The colour-ordered double soft squared amplitude at tree level can be
found in Section 5.3 of Ref.~\cite{Campbell:1997hg} (see also Section
8.1 of Ref.~\cite{GehrmannDeRidder:2005cm}) and it is reported
below~\footnote{We thank Keith Hamilton for providing us with an
  independent derivation of the double soft squared amplitude in
  large-$N_c$.}
\begin{align}
\tilde{w}^{(0)}_{12}(k_a,k_b)&=2\,(2\pi)^4\mu^{4\epsilon}\bigg[\frac{s_{12}^2}{s_{1a}s_{ab2}s_{1ab}s_{b2}}
  +
  \frac{1-\epsilon}{s_{ab}^2}\left(\frac{s_{1a}}{s_{1ab}}+\frac{s_{b2}}{s_{ab2}}-1\right)^2\notag\\
&+\frac{s_{12}}{s_{ab}}\left(\frac{1}{s_{1a}s_{b2}}+\frac{1}{s_{1a}s_{ab2}}+\frac{1}{s_{b2}s_{1ab}}-\frac{4}{s_{1ab}
  s_{ab2}}\right)\bigg]\,,
\end{align}
where the Lorentz invariants $s_{i\dots k}$ indicate the standard
Mandelstam variables.
For later use, it is also convenient to single out the independent
emission contribution and write
\begin{align}
\label{eq:double-soft-real}
\tilde{w}^{(0)}_{12}(k_a,k_b)=\frac{1}{2}w^{(0)}_{12}(k_a) w^{(0)}_{12}(k_b) +  \bar{w}^{(gg)}_{12}(k_a,k_b)\,.
\end{align}
Note, however, that the separation of the independent
contribution is immaterial at the level of the single colour flow,
and only makes physical sense at the level of the sum
$\tilde{w}^{(0)}_{12}(k_a,k_b) + \tilde{w}^{(0)}_{12}(k_b,k_a)$. We
also observe that the \textit{correlated} term $\bar{w}^{(gg)}_{12}$
is not positive definite.
The separation~\eqref{eq:double-soft-real} is useful since the
independent emission contribution is correctly described by the BMS
equation as it simply arises from the iteration of the leading order
squared amplitude. We can therefore focus on the correlated term of
Eq.~\eqref{eq:double-soft-real} in what follows.

The dipole structure of this double-real correction can be read from
the last two terms of
Eq.~\eqref{eq:LL-evolution-kt-iterated}. 
In that equation, the first line corresponds to the usual dipole
structure of the BMS equation, obtained from the replacement
\begin{equation}
\label{eq:replacement_dipole}
\Delta_{1a}(k_{ta}) \Delta_{a2}(k_{ta})\to~G_{1a}[k_{ta};u] G_{a2}[k_{ta};u]\,.
\end{equation}
Similarly, the product of three Sudakov factors in the third and fifth
line encodes the no-emission probability for the three dipoles created
by the emission of $k_a$ and $k_b$ off the initial dipole $\{12\}$.
The dipole structure for the double real correction then corresponds
to replacing 
\begin{align}
\Delta_{1b}(k_{tb}) \Delta_{ba}(k_{tb})\Delta_{a2}(k_{ta})\,\to~ &
                                                                G_{1b}[k_{tb};u] G_{ba}[k_{tb};u]G_{a2}[k_{ta};u]\,,\notag\\
\Delta_{1a}(k_{ta}) \Delta_{ab}(k_{tb})\Delta_{b2}(k_{tb}) \,\to~& G_{1a}[k_{ta};u] G_{ab}[k_{tb};u]G_{b2}[k_{tb};u]\,.
\end{align}
As a last step, we need to upgrade the phase space boundary given
by $\Theta(Q-k_{ta})$ in Eq.~\eqref{eq:LL-evolution-kt-iterated}
which emerges from the strongly ordered limit. Conversely, the
double real correction given by $\bar{w}^{(gg)}_{12}$ describes the
emission of two gluons with commensurate $k_t$ (and energy) with
respect of the emitting $\{12\}$ dipole.
As a consequence, the definition of the resolution scale $Q$ used in
Eq.~\eqref{eq:LLreal} now has to constrain the total energy of
$k_a+k_b$ or equivalently its transverse momentum w.r.t.\ the
$\{12\}$ dipole, in order for the evolution to be collinear safe for
any value of $Q$.
In line with our choice of the dipole transverse momentum as evolution
scale, we therefore perform the replacement
\begin{equation}
\Theta(Q-k_{ta})\to \Theta(Q-k_{t(ab)})\,,
\end{equation}
where we defined
\begin{equation}
\vec{k}_{t(ab)}=\vec{k}_{ta}+ \vec{k}^\prime_{tb}\,,
\end{equation}
and $k_{t(ab)}=|\vec{k}_{t(ab)}|$. We denoted by $\vec{k}^\prime_{tb}$
the transverse momentum of $k_b$ with respect to the $\{12\}$ dipole
and
\begin{equation}
(\vec{k}^\prime_{tb})^2 = 2\frac{(p_1\cdot k_b) (p_2\cdot k_b)}{(p_1\cdot p_2)}\,.
\end{equation}
Notice that $k^\prime_{tb}$ differs significantly from $k_{tb}$ (in
the frame of the emitting dipole of $k_b$) in the limit in which $k_b$
is collinear to $k_a$, and the two have comparable energies. As
already stressed, and we will show shortly, in this limit the real
correction will cancel against the virtual contributions and therefore
this limit is irrelevant for the observable considered here. In this
case, one can only receive a NLL correction from the regime where
$k^\prime_{tb}\sim k_{tb}$.
Subtracting the double counting with the iteration of the BMS
equation~\eqref{eq:LL-evolution-kt-iterated} leads to the following
term to be added to the right hand side of
Eq.~\eqref{eq:LL-evolution-kt}
\begin{align}
\label{eq:tripole}
 \int [d k_a] \int [d k_b]& \abar(k_{ta}) \abar(k_{tb})\frac{\Delta_{12}(Q)}{\Delta_{12}(k_{ta})}\frac{\Delta_{1a}(k_{ta})}{\Delta_{1a}(k_{tb})}
 G_{1b}[k_{tb};u] G_{ba}[k_{tb};u]G_{a2}[k_{ta};u] u(k_a)u(k_b) 
\notag\\
&\times \,\left(\bar{w}^{(gg)}_{12}(k_b,k_a) \Theta(Q-k_{t(ab)})\Theta(k_{ta}-k_{tb}^\prime)\right.\notag\\ & \qquad\qquad-\left.
  w^{(0)}_{12}(k_a) \left(w^{(0)}_{1a}(k_b) -\frac{1}{2} w^{(0)}_{12}(k_b)\right) \Theta(Q-k_{ta})\Theta(k_{ta}-k_{tb})\right) \notag\\
+\int [d k_a] \int [d k_b]& \abar(k_{ta}) \abar(k_{tb}) \frac{\Delta_{12}(Q)}{\Delta_{12}(k_{ta})}\frac{\Delta_{a2}(k_{ta})}{\Delta_{a2}(k_{tb})}
 G_{1a}[k_{ta};u] G_{ab}[k_{tb};u]G_{b2}[k_{tb};u] u(k_a)u(k_b)\notag\\
&\times \left(\bar{w}^{(gg)}_{12}(k_a,k_b) \Theta(Q-k_{t(ab)})\Theta(k_{ta}-k_{tb}^\prime)\right.\\ & \qquad\qquad-\left.w^{(0)}_{12}(k_a) \left(w^{(0)}_{a2}(k_b) -\frac{1}{2} w^{(0)}_{12}(k_b)\right) \Theta(Q-k_{ta})\Theta(k_{ta}-k_{tb})\right) \,,\notag
\end{align}
where the two emissions are ordered in their dipole $k_t$.
In its present form, Eq.~\eqref{eq:tripole} cannot be readily
interpreted as a correction to the kernel of the evolution
equation~\eqref{eq:LL-evolution-kt}, in that it contains the product
of two ratios of Sudakov factors that already indicate an iteration of
the kernel. In order to derive a corresponding term at the level of
the integral equation~\eqref{eq:LL-evolution-kt} we will need to make
some considerations.
As a next step we discuss virtual corrections at NLL order. The BMS
equation already contains virtual corrections in the strongly ordered
soft limit. However, these do not cancel the collinear singularity
present in Eq.~\eqref{eq:tripole} when $k_a$ is collinear to
$k_b$. For this, we need first to introduce the full one-loop
corrections to the evolution kernel in the soft limit.

\paragraph{Virtual corrections.} The second term to consider is the
virtual one-loop correction to the leading order kernel in the r.h.s.\
of Eq.~\eqref{eq:LL-evolution-kt}.
The structure of the virtual corrections can be read off the first
line of Eq.~\eqref{eq:LL-evolution-kt-iterated}. In the large-$N_c$
limit, the emission of $k_a$ generates two adjacent colour dipoles
that emit further radiation incoherently. Therefore, the singularity
structure of virtual corrections to the $\{1a2\}$ configuration will
factorise into the product of virtual corrections to the $\{1a\}$ and
$\{a2\}$ dipoles in this limit (see e.g.~\cite{Gardi:2009qi}).
This factorising structure is already encoded at LL in the first line
of Eq.~\eqref{eq:LL-evolution-kt-iterated}, specifically in the
product $\Delta_{1a}(k_{ta})\Delta_{a2}(k_{ta})$.
However, the LL Sudakov of factors in this product do not contain the
correct NLL singular structure (i.e.\ the single poles), and therefore
we need to supplement the first line of
Eq.~\eqref{eq:LL-evolution-kt-iterated} with an extra correction
factor that accounts for the missing terms.

The one loop corrections to the emission of the soft gluon $k_a$ off
the $\{12\}$ dipole can be written
as~\cite{Catani:2000pi,Angeles-Martinez:2016dph} (given here in the
large-$N_c$ limit)
\begin{equation}
\label{eq:virt-one-loop}
w^{(1)}_{12}(k_a) =w^{(0)}_{12}(k_a) \left[V_{12}^{(1)}(\epsilon)-
    N_c\frac{\alpha_s}{\pi}\,\frac{(4\pi)^\epsilon}{2\epsilon^2}\frac{\Gamma^4(1-\epsilon)\Gamma^3(1+\epsilon)}{\Gamma^2(1-2
  \epsilon)\Gamma(1+2\epsilon)} \left(\frac{\mu^2}{k_{ta}^2}\right)^\epsilon\right] \,,
\end{equation}
where $V_{12}^{(1)}(\epsilon)$ corresponds to one loop virtual
corrections to the dipole $\{12\}$, while the second term encodes the
one loop soft gluon current.
In this context, the quantity $V_{12}^{(1)}(\epsilon)$ is not the
full virtual correction to the Born process, but rather the virtual
correction as predicted by the LL evolution equation.
We derive this by using the unitarity of $G_{12}$
(cf. Eq.~\eqref{eq:unitarityLL}), and expanding the Sudakov
$\Delta(k_{ta})$ at ${\cal O}(\abar)$, obtaining
\begin{equation}
\label{eq:v12-NLL}
  V_{12}^{(1)}(\epsilon)=  - \abar(\mu)\,\int [d k_b] w^{(0)}_{12}(k_b)\Theta(Q-k_{tb})\,.
\end{equation}
The mismatch between Eq.~\eqref{eq:v12-NLL} and the full virtual
correction to the Born process considered here is taken into account
in the hard matching coefficient~${\cal H}_2$ computed in
Sec.~\ref{sec:3jet}.
Moreover, from Eq.~\eqref{eq:virt-one-loop} we also see that choosing
$\mu=k_{ta}$ absorbs all the $k_{ta}$ dependence into the running
coupling.
Renormalising the coupling in the $\overline{\rm MS}$ scheme
(cf. Eq.~\eqref{eq:MS}) allows us to write
\begin{multline}
\label{eq:one-loop}
w^{(1)}_{12}(k_a) \to  - \abar(k_{ta})\, w^{(0)}_{12}(k_a) \\
\times \left[\int [d k_b] \, \,w^{(0)}_{12}(k_b) \Theta(Q-k_{tb}) +
  \frac{e^{\epsilon \gamma_E}}{2\epsilon^2}\frac{\Gamma^4(1-\epsilon)\Gamma^3(1+\epsilon)}{\Gamma^2(1-2
  \epsilon)\Gamma(1+2\epsilon)} + \frac{\bar{\beta}_0}{\epsilon}\right] \,,
\end{multline}
where $\bar{\beta}_0$ is defined by the coefficient of $N_c/\pi$ in
the large-$N_c$ $\beta_0$ given in Eq.~\eqref{eq:beta0_largeN}.

The virtual corrections to the evolution kernel are then obtained from
the first line of Eq.~\eqref{eq:LL-evolution-kt-iterated} via the
replacement
\begin{equation}
\label{eq:matching_virtual}
  \frac{\Delta_{12}(Q)}{\Delta_{12}(k_{ta})}\Delta_{1a}(k_{ta}) \Delta_{a2}(k_{ta})\to~\left(1+\abar(k_{ta})\gamma(k_{a},\epsilon)\right)\,\frac{\Delta_{12}(Q)}{\Delta_{12}(k_{ta})}\Delta_{1a}(k_{ta}) \Delta_{a2}(k_{ta})\,,
\end{equation}
where the function $\gamma(k_a,\epsilon)$ determines the matching
coefficient necessary to obtain the correct virtual corrections to the
emission of a soft gluon.
It is obtained by matching Eq~\eqref{eq:one-loop} and the one-loop
contribution to the r.h.s. of Eq.~\eqref{eq:matching_virtual}.
The expansion of the r.h.s.\ of Eq.~\eqref{eq:matching_virtual} gives
\begin{align}
 w^{(0)}_{12}(k_a)&\left(1+\abar(k_{ta})\gamma(k_{a},\epsilon)\right)\,\frac{\Delta_{12}(Q)}{\Delta_{12}(k_{ta})}\Delta_{1a}(k_{ta})
  \Delta_{a2}(k_{ta}) =  w^{(0)}_{12}(k_a) \left(1+\abar(k_{ta})\gamma(k_{a},\epsilon)\right)\notag\\
& - \abar(k_{ta}) \, w^{(0)}_{12}(k_a)\,\int [d
  k_b]\left[\left(w^{(0)}_{1a}(k_b)+w^{(0)}_{a2}(k_b) -
      w^{(0)}_{12}(k_b) \right)\Theta(k_{ta}-k_{tb}) \right.\notag\\
&\left. + \,w^{(0)}_{12}(k_b)  \Theta(Q-k_{tb})\right] + {\cal O}(\abar^2(k_{ta}))\,.
\end{align}
The coefficient of ${\cal O}(\bar\alpha)$ in the above equation has to
be matched to the one-loop expansion of Eq.~\eqref{eq:one-loop}, from
which we obtain
\begin{align}
\gamma(k_{a},\epsilon) &= -\frac{e^{\epsilon \gamma_E}}{2\epsilon^2}\frac{\Gamma^4(1-\epsilon)\Gamma^3(1+\epsilon)}{\Gamma^2(1-2
  \epsilon)\Gamma(1+2\epsilon)} -
\frac{\bar{\beta}_0}{\epsilon}\notag\\
&+\int [d
  k_b]\left(w^{(0)}_{1a}(k_b)+w^{(0)}_{a2}(k_b) -
      w^{(0)}_{12}(k_b) \right)\Theta(k_{ta}-k_{tb}) \,.
\end{align}
The last integral reads
\begin{equation}
\int [d k_b]\left(w^{(0)}_{1a}(k_b)+w^{(0)}_{a2}(k_b) -
      w^{(0)}_{12}(k_b) \right)\Theta(k_{ta}-k_{tb})  = \left(\frac{\mu^2}{k_{ta}^2}\right)^{\epsilon}\left(\frac{1}{2\epsilon^2}-\frac{\pi^2}{24}+{\cal O}(\epsilon)\right)\,,
\end{equation}
which gives (setting $\mu=k_{ta}$)
\begin{equation}
\label{eq:gammadef}
\gamma(k_{a},\epsilon) = \gamma(\epsilon)= -\frac{11}{12\epsilon}+\frac{\pi^2}{6}+{\cal O}(\epsilon)\,.
\end{equation}

\paragraph{Cancellation of collinear singularities.} 
We now combine the real and virtual corrections obtained above into a
single integral equation in $d=4-2 \epsilon$ dimensions.
We want to achieve a manifest cancellation of soft and collinear
singularities between real and virtual contributions independently
of the precise form of the source $u(k)$. For this, we make use of
the fact that, as stressed earlier, the double-real corrections of
Eq.~\eqref{eq:tripole} only contribute in the unordered regime where
$\omega_a\sim \omega_b$.  For the observables considered in this
article, this also implies $k_{ta}\sim k_{tb}\sim k_{t(ab)}$.
To see this, we observe that $k_{ta} \sim k_{tb}$ only if
$\omega_a\sim \omega_b$ and the two gluons are not collinear to one
another. Away from this configuration, when the two gluons become
collinear, one can end up in a situation with $\omega_a\sim \omega_b$
but the dipole transverse momenta are strongly ordered, i.e.\
$k_{ta}\gg k_{tb}$. This configuration however only contributes to the
observable if both gluons are inside the rapidity slice, hence at wide
angles w.r.t.\ the $\{12\}$ dipole, and the corresponding collinear
singularity exactly cancels against the one in
$\gamma(\epsilon)$~\eqref{eq:gammadef} at all orders in $\bar\alpha$,
leaving behind only finite contribution without a logarithmic
enhancement (i.e.\ NNLL). Conversely, for an observable sensitive also
to collinear radiation (e.g. the light-hemisphere mass), this
configuration can occur with both soft gluons being simultaneously
collinear to one of the $\{12\}$ dipole ends, and to each other. The
singularity structure of this configuration would result in an extra
logarithmic enhancement and the argument made above would not hold.
This extra logarithmic enhancement can be however resummed by means of
standard techniques used for global observables, and the formalism
presented here can still be used for the calculation of the non-global
contributions.
We can therefore perform a first-order Taylor expansion and make the
following approximation in the double real corrections given in
Eq.~\eqref{eq:tripole}
\begin{equation}
\abar(k_{ta})
\abar(k_{tb})\frac{\Delta_{12}(Q)}{\Delta_{12}(k_{ta})}\frac{\Delta_{1a}(k_{ta})}{\Delta_{1a}(k_{tb})}\sim \abar^2(k_{ta})
\frac{\Delta_{12}(Q)}{\Delta_{12}(k_{ta})}\sim \abar^2(k_{t(ab)})
\frac{\Delta_{12}(Q)}{\Delta_{12}(k_{t(ab)})}\,,
\end{equation}
(and similarly for the second term corresponding to the alternative
colour flow) where we systematically neglected corrections of order
\begin{equation}
\label{eq:logsApprox}
{\cal O}\left(\alpha_s\ln k_{ta}/k_{t(ab)}\right)\sim {\cal
 O}\left(\alpha_s\ln k_{ta}/k_{tb}\right) \sim {\cal
 O}\left(\alpha_s\right) \sim {\cal O}\left({\rm NNLL}\right)\,,
\end{equation}
in the limit of soft-wide-angle radiation. This is because the
logarithm of the ratios $k_{ta}/k_{t(ab)}$ and $k_{ta}/k_{tb}$ are not
large in the region where Eq.~\eqref{eq:tripole} is non vanishing, and
the quantity~\eqref{eq:logsApprox} only gives a contribution of
${\cal O}(1)$ upon integration over the phase space of $k_a$ and
$k_b$, that is a NNLL correction.
We can write the NLL integral equation in $d$ dimensions as
\begin{align}
\label{eq:NLL-evolution-dimreg}
&G_{12}[Q;u] =\Delta_{12}(Q)+ \int [d k_a] \abar(k_{ta}) w^{(0)}_{12}(k_a)
\,\bigg(1 +\abar(k_{ta})\,\gamma(\epsilon)\bigg)\\
&\times\frac{\Delta_{12}(Q)}{\Delta_{12}(k_{ta})}G_{1a}[k_{ta};u]
  G_{a2}[k_{ta};u] u(k_a)\Theta(Q-k_{ta})\notag\\
&+\int [d k_a] \int [d k_b] \abar^2(k_{t(ab)})
  \frac{\Delta_{12}(Q)}{\Delta_{12}(k_{t(ab)})}u(k_a)u(k_b)\Theta(Q-k_{t(ab)}) \Theta(k_{ta}-k_{tb}^\prime)\notag\\
&\times\left[ \bar{w}^{(gg)}_{12}(k_b,k_a)  G_{1b}[k_{t(ab)};u]
  G_{ba}[k_{t(ab)};u]G_{a2}[k_{t(ab)};u]\right. \notag\\
&\left.\hspace{3cm}+ \bar{w}^{(gg)}_{12}(k_a,k_b) G_{1a}[k_{t(ab)};u] G_{ab}[k_{t(ab)};u]G_{b2}[k_{t(ab)};u]\right] \notag\\
&-\int [d k_a] \int [d k_b] \abar^2(k_{ta})
  \frac{\Delta_{12}(Q)}{\Delta_{12}(k_{ta})}u(k_a)u(k_b)
  \Theta(Q-k_{ta}) \Theta(k_{ta}-k_{tb})\notag\\
&\times\left[w^{(0)}_{12}(k_a) \left(w^{(0)}_{1a}(k_b) -\frac{1}{2}
  w^{(0)}_{12}(k_b)\right)  G_{1b}[k_{ta};u]
  G_{ba}[k_{ta};u]G_{a2}[k_{ta};u] \right.\notag\\
&\left.\hspace{3cm}+ w^{(0)}_{12}(k_a) \left(w^{(0)}_{a2}(k_b) -\frac{1}{2} w^{(0)}_{12}(k_b)\right) G_{1a}[k_{ta};u] G_{ab}[k_{ta};u]G_{b2}[k_{ta};u] \right]\notag\,,
\end{align}
where we have also Taylor expanded the scale of $G_{12}$ in the double
real correction following the same argument as in
Eq.~\eqref{eq:logsApprox}.
We will shortly discuss a way to make the subtraction of IRC
divergences manifest and eventually to take the limit
$\epsilon \to 0$.

\paragraph{The NLL Sudakov form factor.} As a consistency check of the
above equation, we can use Eq.~\eqref{eq:NLL-evolution-dimreg} to
derive the NLL soft anomalous dimension that enters the evolution
equation of the soft corrections to the Sudakov form factor given in
Ref.~\cite{Magnea:2000ss}.
We start by setting the source $u$ to 1 (hence $G[1]=1$) in
Eq.~\eqref{eq:NLL-evolution-dimreg} and obtain
\begin{align}
1 =&\Delta_{12}(Q)+ \int [d k_a] \abar(k_{ta}) w^{(0)}_{12}(k_a)
\,\bigg(1+ \abar(k_{ta})\,\gamma(\epsilon)\bigg)\frac{\Delta_{12}(Q)}{\Delta_{12}(k_{ta})}\Theta(Q-k_{ta})\notag\\
&+\int [d k_a] \int [d k_b] \abar^2(k_{t(ab)})
                                                                                                                     \frac{\Delta_{12}(Q)}{\Delta_{12}(k_{t(ab)})}\Theta(Q-k_{t(ab)})\Theta(k_{ta}-k_{tb}^\prime)\notag\\
   &\times\left[ \bar{w}^{(gg)}_{12}(k_b,k_a)  + \bar{w}^{(gg)}_{12}(k_a,k_b) \right] \notag
  \\
&-\int [d k_a] \int [d k_b] \abar^2(k_{ta})
  \frac{\Delta_{12}(Q)}{\Delta_{12}(k_{ta})}\Theta(Q-k_{ta}) \Theta(k_{ta}-k_{tb})\notag\\
&\times w^{(0)}_{12}(k_a) \left(w^{(0)}_{1a}(k_b)+w^{(0)}_{a2}(k_b) - w^{(0)}_{12}(k_b)\right)\,.
\end{align}
We now introduce a $d$-dimensional massless momentum $k$ such that
$k_t\coloneqq|\vec{k}_t|=k_{t(ab)}$ and its  rapidity $\eta$
with respect to the emitting dipole $\{12\}$ is that of the $k_a+k_b$
system $\eta_{(ab)}$, this is defined by the corresponding kinematic
map\footnote{An analogous map has been considered in the context of
  the resummation of global observables in
  refs.~\cite{Banfi:2014sua,Banfi:2018mcq,Bauer:2019bsp}.}
\begin{equation}
\label{eq:map_ab}
{\mathbb P} : \{k_a,k_b\}\rightarrow k_{(ab)} = 
 \left(k_{t(ab)}\cosh \eta_{(ab)},\vec{k}_{t(ab)},k_{t(ab)}\sinh \eta_{(ab)}\right)
\,,
\end{equation}
where the three-vectors are taken in the rest frame of the $\{12\}$
dipole.
We recast the above equation as
\begin{align}
\label{eq:sudakov_NLL_eq}
  1 =&\Delta_{12}(Q)+ \int [d k_a] \abar(k_{ta}) w^{(0)}_{12}(k_a)
       \,\bigg(1+ \abar(k_{ta})\,\gamma(\epsilon)\bigg)\frac{\Delta_{12}(Q)}{\Delta_{12}(k_{ta})}\Theta(Q-k_{ta})\notag\\
     &+\int [d k]\, \bar{\delta}(k-k_a-k_b) \abar^2(k_{t})
       \frac{\Delta_{12}(Q)}{\Delta_{12}(k_{t})}\Theta(Q-k_{t})\notag\\
     &\times\int [d k_a] \int [d k_b] 
       \Theta(k_{ta}-k_{tb}^\prime)
       \left( \bar{w}^{(gg)}_{12}(k_b,k_a)  + \bar{w}^{(gg)}_{12}(k_a,k_b) \right)\notag \\
     &-\int [d k_a] \int [d k_b] \abar^2(k_{ta})
       \frac{\Delta_{12}(Q)}{\Delta_{12}(k_{ta})}\Theta(Q-k_{ta}) \Theta(k_{ta}-k_{tb})\notag\\
     &\times w^{(0)}_{12}(k_a)  \left(w^{(0)}_{1a}(k_b)
       +w^{(0)}_{a2}(k_b) -
       w^{(0)}_{12}(k_b)\right)\,,
\end{align}
where we defined
\begin{equation}
\label{eq:delta3}
\bar{\delta}(k-k_a-k_b)\coloneqq 2 (2\pi)^{3-2\epsilon}\delta^{(2-2\epsilon)}(\vec{k}_t-\vec{k}_{ta}-\vec{k}^\prime_{tb})\delta(\eta-\eta_{(ab)}) \,.
\end{equation}
We now divide Eq.~\eqref{eq:sudakov_NLL_eq} by $\Delta_{12}(Q)$ and
take the derivative with respect to $\ln Q$, obtaining the differential
evolution equation
\begin{align}
\label{eq:sudakov_RGE_gamma}
&\frac{d\ln\Delta_{12}(Q)}{d\ln Q} =- \int [d k] Q\delta(Q-k_{t})\bigg\{\abar(k_{t}) w^{(0)}_{12}(k)
\,\bigg(1+\abar(k_{t})\,\gamma(\epsilon)\bigg)\\
&+ \abar^2(k_{t})
  \int [d k_a]
  \int [d k_b] \bar{\delta}(k-k_a-k_b)\bigg[
  \,\bar{w}^{(gg)}_{12}(k_b,k_a)+\bar{w}^{(gg)}_{12}(k_a,k_b)\bigg]\Theta(k_{ta}-k_{tb}^\prime)\notag\\
& -\abar^2(k_{t}) \int [d k_a]\int [d k_b]\bar{\delta}(k-k_a) \left[w^{(0)}_{12}(k_a)
  \left(w^{(0)}_{1a}(k_b) + w^{(0)}_{a2}(k_b)-
  w^{(0)}_{12}(k_b)\right) \right]\Theta(k_{ta}-k_{tb})\bigg\}\,.\notag
\end{align}
In the second line of Eq.~\eqref{eq:sudakov_RGE_gamma}, it is
convenient to parametrise the double real phase space as
in~\cite{Dokshitzer:1997iz} (see also Ref.~\cite{Banfi:2018mcq} for
the $d$-dimensional case)
\begin{equation}
\label{eq:ps2}
[dk_a][dk_b]= [dk] \frac{dz
  [z(1-z)]^{-\epsilon}}{(4\pi)^2}\frac{dm_{ab}^2}{m_{ab}^{2\epsilon}} \frac{d\Omega_{2-2\epsilon}}{(2\pi)^{1-2\epsilon}}\,,
\end{equation}
where $[dk]$ is given in Eq.~\eqref{eq:phase-space-kt},
$m_{ab}^2=(k_a+k_b)^2 \in [0,\infty)$, and $z \in [0,1]$ is the
fraction of one of the light cone components of $k$ with respect to
the $\{12\}$ dipole carried by one of the daughters $k_a$ or $k_b$. For
instance if we adopt the following Sudakov parametrisation for the
four momenta
\begin{align}
k_{a/b}^\mu &= z_{a/b}^{(1)}\,p_1^\mu+z_{a/b}^{(2)}\,p_2^\mu +
              \kappa_{{a/b}}^\mu\notag\\
k^\mu &= z^{(1)}\,p_1^\mu+z^{(2)}\,p_2^\mu + \kappa^\mu\,,
\end{align}
with $\kappa_i$ being a space-like transverse vector describing the
transverse momentum of $k_i$ w.r.t.\ the $\{12\}$ dipole, we can define
$z$ as
\begin{align}
z_{a}^{(1)} = z \,z^{(1)}\,\quad z_{b}^{(1)} = (1-z) \,z^{(1)}\,.
\end{align}
Finally, $d\Omega_{2-2\epsilon}$ is the angular phase space for the
vector $\vec q \coloneqq \frac{\vec k_{ta}}{z}-\frac{\vec k^\prime_{tb}}{1-z}$
with respect to $\vec k_t$, given by
\begin{equation}
  \label{eq:azimuthal-ps}
  \frac{d\Omega_{2-2\epsilon}}{(2\pi)^{1-2\epsilon}} =
  \frac{(4\pi)^{\epsilon}}{\sqrt{\pi}\Gamma(\frac{1}{2}-\epsilon)}
  d\phi \,(\sin^2\phi)^{-\epsilon}\,, \qquad \phi\in [0,\pi]\,.
\end{equation}
Since $k$ is massless, we can integrate inclusively over $m_{ab}$ in
the second line of Eq.~\eqref{eq:sudakov_RGE_gamma}, and find
\begin{align}
\label{eq:correl_eps}
&\int [d k_a]
  \int [d k_b] \,\bar{\delta}(k-k_a-k_b)\bigg[
  \,\bar{w}^{(gg)}_{12}(k_b,k_a)+\bar{w}^{(gg)}_{12}(k_a,k_b)\bigg]\Theta(k_{ta}-k_{tb}^\prime)\notag\\
&\qquad  =
  w^{(0)}_{12}(k)\left(\frac{\mu^2}{k_{t}^2}\right)^{\epsilon}\left(\frac{1}{2\epsilon^2}+\frac{11}{12\epsilon}+\frac{67}{36}-\frac{7}{24}\pi^2+{\cal
  O}(\epsilon)\right)\,,\\
\label{eq:correl_eps_SO}
&\int [d k_a]\int [d k_b]\,\bar{\delta}(k-k_a) \left[w^{(0)}_{12}(k_a)
  \left(w^{(0)}_{1a}(k_b) + w^{(0)}_{a2}(k_b)-
  w^{(0)}_{12}(k_b)\right) \right]\Theta(k_{ta}-k_{tb}) \notag\\
&\qquad=  w^{(0)}_{12}(k)\left(\frac{\mu^2}{k_{t}^2}\right)^{\epsilon}\left(\frac{1}{2\epsilon^2}-\frac{\pi^2}{24}+{\cal
 O}(\epsilon)\right)\,.
\end{align}
This leads to the well known evolution equation for the soft radiative
corrections to the Sudakov form factor (see
e.g.~\cite{Laenen:2000ij,Magnea:2000ss,Dixon:2008gr}), which can be
expressed as~\cite{Laenen:2000ij,Banfi:2018mcq}
\begin{equation}
\label{eq:sudakov_RGE_NLL}
\frac{d\ln\Delta_{12}(Q)}{d\ln Q} = - \int [d k] Q\delta(Q-k_{t})
\abar(k_{t}) w^{(0)}_{12}(k) \left(1+\abar(k_t)
  \bar{K}^{(1)}+{\cal O}(\abar^2(k_t))\right)\,,
\end{equation}
where 
\begin{equation}
\bar{K}^{(1)}=\frac{67}{36}-\frac{\pi^2}{12}
\end{equation}
is the coefficient of $N_c$ of the large-$N_c$ two-loop cusp anomalous
dimension in units of $\alpha_s/\pi$.
One can directly use this result and
Eq.~\eqref{eq:NLL-evolution-dimreg} to derive a differential equation
for $G_{12}$ in $d$ dimensions. However, since we are ultimately
interested in a numerical evaluation of its solution, we first discuss
how to take the limit $\epsilon\to 0$.

\subsection{The NLL differential equation}
In order to take the $\epsilon\to 0$ limit we wish to make the
cancellation of infrared singularities in
Eq.~\eqref{eq:NLL-evolution-dimreg} manifest. This is crucial to solve
the evolution equation numerically for different infrared safe
observables. To this end we introduce the counterterm
\begin{align}
\label{eq:counter-term}
&\int [d k_a] \int [d k_b] \abar^2(k_{t(ab)})
                          \frac{\Delta_{12}(Q)}{\Delta_{12}(k_{t(ab)})}u(k_{(ab)})\Theta(Q-k_{t(ab)})\Theta(k_{ta}-k_{tb}^\prime)\notag\\
&\qquad \times\left[ \bar{w}^{(gg)}_{12}(k_b,k_a)  +
  \bar{w}^{(gg)}_{12}(k_a,k_b) \right] G_{1(ab)}[k_{t(ab)};u] G_{(ab)2}[k_{t(ab)};u]\notag\\
&-\int [d k_a] \int [d k_b] \abar^2(k_{ta})
  \frac{\Delta_{12}(Q)}{\Delta_{12}(k_{ta})}u(k_a)
  \Theta(Q-k_{ta}) \Theta(k_{ta}-k_{tb})\notag\\
&\qquad\times\left[w^{(0)}_{12}(k_a) \left(w^{(0)}_{1a}(k_b) +w^{(0)}_{12}(k_a)-
  w^{(0)}_{12}(k_b)\right) \right] G_{1a}[k_{ta};u]
G_{a2}[k_{ta};u]\,.
\end{align}
which we add and subtract to Eq.~\eqref{eq:NLL-evolution-dimreg}.  The
main difference between Eq.~\eqref{eq:counter-term} and the double
real correction~\eqref{eq:tripole} is the source evaluated on the
\textit{massless} momentum $k_{(ab)}$ defined in
Eq.~\eqref{eq:map_ab}.
This definition, owing to the collinear safety of source $u(k)$ in
Eq.~\eqref{eq:counter-term}, guarantees that the collinear singularity
is consistently cancelled in the difference between the double-real
contribution to Eq.~\eqref{eq:NLL-evolution-dimreg} and the
counterterm~\eqref{eq:counter-term}.
In general, one can replace the momentum $k_{(ab)}$ with the result of
any kinematic map ${\mathbb P} : \{k_a,k_b\}\rightarrow k_{(ab)}$ such
that in the limit where $k_a$ and $k_b$ are collinear, the transverse
momentum and  rapidity of the
$k_{(ab)}$ momentum coincide with those of the $k_a+k_b$ system with
respect to the $\{12\}$ dipole. The choice of the kinematic map used
in the subtraction is arbitrary, and here we adopt the {\it massless
  projection} of Eq.~\eqref{eq:map_ab} since it makes the computation
of the integrated counterterm simple.

A second important difference between Eq.~\eqref{eq:counter-term} and
Eq.~\eqref{eq:tripole} is that the generating functionals $G_{1(ab)}$
and $G_{(ab)2}$ now depend on the angle of the momentum $k_{(ab)}$
with respect to the $\{12\}$ dipole defined by the above
projection~\eqref{eq:map_ab}.
We now deal with the integral of the
counterterm~\eqref{eq:counter-term}, which is instead combined with
the real-virtual corrections in the first two lines of
Eq.~\eqref{eq:NLL-evolution-dimreg}.
Given the massless nature of $k_{(ab)}$, we can adopt the
parametrisation~\eqref{eq:ps2} and integrate as in
Eq.~\eqref{eq:correl_eps}. The integrated counterterm then cancels
the $\epsilon$ poles of $\gamma(\epsilon)$ in
Eq.~\eqref{eq:NLL-evolution-dimreg}, and will later allow us to take
the limit $\epsilon\to 0$ at the integrand level.
One can finally derive the corresponding integro-differential
equation, and bring it in the form given in
Eq.~\eqref{eq:NLL-evolution-kt-diff-symbolic}.
As done for the evolution equation for the NLL Sudakov factor, we
start by dividing Eq.~\eqref{eq:NLL-evolution-dimreg} supplemented
with the counterterm introduced above by $\Delta_{12}(Q)$ and then we
take the derivative with respect to $\ln Q$. Using the evolution
equation for the Sudakov factor~\eqref{eq:sudakov_RGE_NLL} one finds
that the dependence on the Sudakov factors entirely drops out and we
obtain the NLL evolution equation
\begin{align}
\label{eq:NLL-evolution-kt-diff}
Q \partial_Q G_{12}[Q;u] &={\mathbb K}^{\rm NLL}[G[Q,u],u]\notag\\
&\coloneqq {\mathbb K}^{\rm RV+VV}[G[Q,u],u]+{\mathbb K}^{\rm RR}[G[Q,u],u]-{\mathbb K}^{\rm DC}[G[Q,u],u]\,.
\end{align}
where we have defined
\begin{align}
\label{eq:Uvirtual}
{\mathbb K}^{\rm RV+VV}&[G[Q,u],u] \coloneqq \int [d k_a] \abar(Q) w^{(0)}_{12}(k_a)
\,\bigg(1 +\abar(Q)\,\bar{K}^{(1)}\bigg)\\
&\hspace{1cm}\times\left(G_{1a}[Q;u]
  G_{a2}[Q;u] u(k_a) - G_{12}[Q;u]\right)Q\delta(Q-k_{ta})\notag\,,
\end{align}
as the kernel correction due to the virtual and subtracted
real-virtual corrections,
\begin{align}
\label{eq:Ureal}
{\mathbb K}^{\rm RR}&[G[Q,u],u] \coloneqq  \int [d k_{a}]\int
                      [d k_{b}] \,\abar^2(Q)Q\delta(Q-k_{t(ab)}) \Theta(k_{ta}-k_{tb}^\prime)\\
&\times\left[ \bar{w}^{(gg)}_{12}(k_b,k_a)  G_{1b}[Q;u]
  G_{ba}[Q;u]G_{a2}[Q;u]u(k_a)u(k_b)\right. \notag\\
&\left.\hspace{1cm}+ \bar{w}^{(gg)}_{12}(k_a,k_b) G_{1a}[Q;u]
  G_{ab}[Q;u]G_{b2}[Q;u]u(k_a)u(k_b)\right.\notag\\
&\left. \hspace{2cm} - \left(\bar{w}^{(gg)}_{12}(k_b,k_a) +\bar{w}^{(gg)}_{12}(k_a,k_b)\right)G_{1(ab)}[Q;u]
  G_{(ab)2}[Q;u]u(k_{(ab)})\right]\notag\,,
\end{align}
for the double real corrections, and finally
\begin{align}
\label{eq:Udc}
{\mathbb K}^{\rm DC}&[G[Q,u],u] \coloneqq \int [d k_a] \int [d k_b] \abar^2(Q) Q\delta(Q-k_{ta}) \Theta(k_{ta}-k_{tb})\\
&\times\biggl[w^{(0)}_{12}(k_a) \left(w^{(0)}_{1a}(k_b) -\frac{1}{2}
  w^{(0)}_{12}(k_b)\right)  G_{1b}[Q;u]
  G_{ba}[Q;u]G_{a2}[Q;u]u(k_a)u(k_b) \notag\\
&\hspace{1cm}+ w^{(0)}_{12}(k_a) \left(w^{(0)}_{a2}(k_b)
  -\frac{1}{2} w^{(0)}_{12}(k_b)\right) G_{1a}[Q;u]
  G_{ab}[Q;u]G_{b2}[Q;u] u(k_a)u(k_b)\notag\\
&\hspace{2cm}-w^{(0)}_{12}(k_a) \left(w^{(0)}_{1a}(k_b)  +w^{(0)}_{a2}(k_b)- w^{(0)}_{12}(k_b)\right) G_{1a}[Q;u]
  G_{a2}[Q;u]u(k_a)\biggr]\notag\,,
\end{align}
for the subtraction of the double counting with the iteration of the
LL evolution kernel.  In taking the four-dimensional limit in Eqs.~\eqref{eq:Ureal} and~\eqref{eq:Udc}, while keeping NLL contributions only, we implicitly neglect NNLL configurations in which $k_a$ and $k_b$ are both inside the rapidity slice. This is also crucial to guarantee the collinear safety of Eq.~\eqref{eq:Udc}.
Finally, we observe that since the observables under consideration are
sensitive only to radiation at small rapidities, the exact rapidity
bound~\eqref{eq:rapidity-bound} in the phase space integral is
irrelevant. Therefore, it can be relaxed and set as
\begin{equation}
\label{eq:rapinf}
-\infty < \eta_{a/b}<+\infty
\end{equation}
in the differential equations~\eqref{eq:LL-evolution-kt-diff-symbolic}
and~\eqref{eq:NLL-evolution-kt-diff}, which are both collinear safe
and thus well defined since outside the slice there is a complete
cancellation between real and virtual corrections. On the other hand,
in the corresponding integral equations, virtual corrections are
encoded in the Sudakov form factors. Therefore, one cannot use
Eq.~\eqref{eq:rapinf} since real and virtual terms require a finite
rapidity upper bound in order to be separately well defined. In this
case the initial rapidity bound~\eqref{eq:rapidity-bound} must be
retained although the insensitivity of the observable to the large
rapidity region ensures that the dependence on this cancels out
eventually.
The boundary conditions to Eq.~\eqref{eq:NLL-evolution-kt-diff} in $d$
dimensions are given in Eq.~\eqref{eq:initial-cond}.

\subsection{Limit to $d=4$ and boundary conditions}
\label{sec:4D}
We now briefly comment on taking the $\epsilon\to 0$ limit of
Eqs.~\eqref{eq:LL-evolution-kt-diff-symbolic}
and~\eqref{eq:NLL-evolution-kt-diff}.
The right-hand side of both equations is now finite and well defined
in four dimensions as long as the evolution scale is larger than
$\Lambda_{\rm QCD}$, so that the limit $\epsilon \to 0$ can be
directly taken at the integrand level.
However, the boundary condition given in Eq.~\eqref{eq:initial-cond}
is not well defined in this limit because of the Landau singularity.
One has to supplement the evolution equations for $G_{12}$ with some
non-perturbative modelling that allows the computation at very small
scales.
A simple prescription in the context of a numerical calculation is to
introduce a freezing of the coupling at some small scale
$Q_0 > \Lambda_{\rm QCD}$ such that
\begin{equation}
\label{eq:initial-cond-4D-alt}
\alpha_s(k) = \alpha_s(Q_0),\qquad k \leq Q_0\,.
\end{equation}
An alternative prescription is to require that $G_{12}$ is kept
constant below the freezing point $Q_0$, resulting in the boundary
condition
\begin{equation}
\label{eq:initial-cond-4D}
G_{12}[Q;u] = 1~{\rm for}~ Q \leq Q_0\,,
\end{equation}
while the unitarity condition $G_{12}[Q;1]=1$ is unchanged. The
physical picture corresponding to Eq.~\eqref{eq:initial-cond-4D} is
that, below the resolution scale $Q_0$, real radiation cancels
exactly virtual corrections by virtue of unitarity.
This of course requires that the freezing scale $Q_0 \ll v$, which is
the typical scale of soft radiation inside the observed rapidity
gap. If this condition is satisfied, the dependence on the infrared
scale $Q_0$ becomes numerically negligible.
In their four-dimensional formulation,
Eqs.~\eqref{eq:LL-evolution-kt-diff}
and~\eqref{eq:NLL-evolution-kt-diff} can be evaluated numerically with
the boundary conditions~\eqref{eq:initial-cond-4D-alt},
or~\eqref{eq:initial-cond-4D}, either via discretisation techniques or
by Monte Carlo methods. We will address the numerical solution in a
forthcoming publication.

\section{Computation of the one-loop hard matching factors}
\label{sec:3jet}
We now discuss the hard factors in Eq.~\eqref{eq:master}. These
account for the contribution of hard radiation that propagates outside
of the observed region of phase space (rapidity slice in our case).
Below we will give a definition that allows for a numerical
implementation of the formulae derived in this article.

To extract the matching coefficients ${\cal H}_2$ and ${\cal H}_3$, it
is instructive to see how they arise from an explicit NLO calculation
in the case of $e^+e^-\to $ jets. We recall that here we use a
flavour-based labelling of momenta, where $p_1$ and $p_2$ are the
momenta of the quark and the antiquark, and $p_3$ the momentum of an
additional hard gluon. The ${\cal O}(\alpha_s)$ virtual contribution
to the cumulative cross section reads (in the $\overline{\rm MS}$
scheme and with $\alpha_s=\alpha_s(\mu)$)
\begin{equation}
\label{eq:Sigmav-2jet}
\Sigma^{{\rm virt.}}(v) =
C_F\frac{\alpha_s}{2\pi}\left(\frac{\mu^2}{s}\right)^{\epsilon}\frac{e^{\gamma_E\epsilon}}{\Gamma(1-\epsilon)}\left(-\frac{2}{\epsilon^2}
  -\frac{3}{\epsilon} +\pi^2 -8\right) S_2(v)\,.
\end{equation}
To obtain the real corrections, and in order to avoid introducing
additional abstract notation, we explicitly parametrise the phase
space in terms of the energy fraction of the gluon $x\coloneqq 2E_g/\sqrt{s}$
and the cosine of the angle between the gluon and the quark
$y\coloneqq\cos\theta_{qg}$, obtaining 
\begin{align}
  \label{eq:Sigmav-3jet}
 & \Sigma^{{\rm real}}(v) =
  2C_F\frac{\alpha_s}{2\pi}\left(\frac{\mu^2}{s}\right)^{\epsilon}\frac{e^{\gamma_E\epsilon}}{\Gamma(1-\epsilon)}\int_0^1d
                                      x \frac{x^{-1-2
                                      \epsilon}}{(1-x)^{2
                                      \epsilon}}\int_{-1}^1 d y\,  \frac{(1-y)^{-1-\epsilon} (1+y)^{-1-\epsilon}}{ (2-x
   (1-y))^{2 -2\epsilon} }\notag\\
&\times\left(8-\epsilon x^2 (2-x (1-y))^2-(2-x) x \left((x-2) x (1-y)^2-4 y+8\right)\right)\Theta_{\rm out}(k) S_{3}(v)\,.
\end{align}
The three directions $\vec{n}_i$ ($i=1,2,3$) in $S_3$ correspond to the
directions of the quark, antiquark and gluon that are entirely
specified by the variables $x$ and $y$.
The phase space constraint $\Theta_{\rm out}(k)$ ensures that no hard
parton is inside the observed rapidity slice, so that the observable
is zero in a pure three-parton configuration. It can be entirely
parametrised in terms of the kinematics of the gluon $k$.
Configurations with a hard parton inside the slice would just
correspond to power corrections and can be accounted for at fixed
perturbative order through standard matching procedures.
Upon integration over $x$ and $y$, Eq.~\eqref{eq:Sigmav-3jet} develops
double and single poles in $\epsilon$ that exactly cancel against
those in Eq.~\eqref{eq:Sigmav-2jet}.

In order to derive the expressions for ${\cal H}_2$ and ${\cal H}_3$,
we now need to subtract the double counting with the real and virtual
corrections to the evolution equation at ${\cal O}(\alpha_s)$. These
can be derived from the results of the previous section. The virtual
corrections are given in Eq.~\eqref{eq:v12-NLL} and at
${\cal O}(\alpha_s)$ read (we set the upper bound of the evolution
scale $Q=\sqrt{s}$)
\begin{align}
\Sigma^{{\rm virt.}}_{\rm soft}(v) &=   -4C_F\frac{\alpha_s}{2\pi}\mu^{2\epsilon}\frac{e^{\gamma_E\epsilon}}{\Gamma(1-\epsilon)}\int_{0}^{\sqrt{s}}\frac{d
    k_t}{k_t^{1+2\epsilon}}\int_{\ln (k_t/\sqrt{s})}^{\ln (\sqrt{s}/k_t)}d \eta \,S_2(v)\notag\\
& \qquad = C_F\frac{\alpha_s}{2\pi}\left(\frac{\mu^2}{s}\right)^{\epsilon}\frac{e^{\gamma_E\epsilon}}{\Gamma(1-\epsilon)}\left(-\frac{2}{\epsilon^2}\right) S_2(v)\,.
\end{align}
Similarly, the real corrections can be obtained from the first
iteration of the evolution equation~\eqref{eq:LL-evolution-kt},
retaining only the contribution in which the soft gluon $k$ is outside
the slice, obtaining
\begin{align}
\label{eq:Sigmav-3jet-soft}
\Sigma^{{\rm real}}_{\rm soft}(v) &= 4C_F\frac{\alpha_s}{2\pi}\mu^{2\epsilon}\frac{e^{\gamma_E\epsilon}}{\Gamma(1-\epsilon)}\int_{0}^{\sqrt{s}}\frac{d
    k_t}{k_t^{1+2\epsilon}}\int_{\ln (k_t/\sqrt{s})}^{\ln (\sqrt{s}/k_t)}d \eta
                                     \,S^{\rm soft}_3(v) \Theta^{\rm soft}_{\rm out}(k)\,,
\end{align}
where $S^{\rm soft}_3$ indicates that the gluon $k$ defining the
third direction $\vec{n}_3$ is soft, so no recoil is present in the
event kinematics.
In the above equation $\Theta^{\rm soft}_{\rm out}(k)$ ensures again that
no partons are present inside the slice at ${\cal O}(\alpha_s)$. Its
expression differs from that of $\Theta_{\rm out}(k)$ in that no
kinematic recoil is present in the soft limit and therefore the thrust
axis is aligned with the direction of the quark (antiquark).

From there, we can compute the hard contribution to the virtual
corrections as
\begin{equation}
\Sigma^{{\rm virt.}}(v)-\Sigma^{{\rm virt.}}_{\rm soft}(v)=C_F\frac{\alpha_s}{2\pi}\left(\frac{\mu^2}{s}\right)^{\epsilon}\frac{e^{\gamma_E\epsilon}}{\Gamma(1-\epsilon)}\left(-\frac{3}{\epsilon} +\pi^2 -8\right) S_2(v)\,,
\end{equation}
which, as expected, contains only a single pole of hard-collinear nature.
We now consider the difference between the real corrections given in
Eqs.~\eqref{eq:Sigmav-3jet},~\eqref{eq:Sigmav-3jet-soft}. Before
taking the difference, we observe that the phase space for the
emission of a soft gluon in Eq.~\eqref{eq:Sigmav-3jet-soft} is larger
than the one imposed by momentum conservation in
Eq.~\eqref{eq:Sigmav-3jet}. This difference comes from having expanded
consistently the rapidity boundary as in Eq.~\eqref{eq:rapidity-bound}
as well as from taking $k_t \leq \sqrt{s}$ rather than $\sqrt{s}/2$ as
required by energy conservation.
We can therefore recast Eq.~\eqref{eq:Sigmav-3jet-soft} as the sum of
a term with the same phase-space limits as Eq.~\eqref{eq:Sigmav-3jet}
and a remainder as
\begin{align}
\label{eq:Sigmav-3jet-soft-recast}
\Sigma^{{\rm real}}_{\rm soft}(v) &= 4C_F\frac{\alpha_s}{2\pi}\mu^{2\epsilon}\frac{e^{\gamma_E\epsilon}}{\Gamma(1-\epsilon)}\bigg[\int_{0}^{\frac{\sqrt{s}}{2}}\frac{d
    k_t}{k_t^{1+2\epsilon}}\int_{-\cosh^{-1}\frac{\sqrt{s}}{2k_t}}^{\cosh^{-1}\frac{\sqrt{s}}{2k_t}}d
                                     \eta + \int_{0}^{\sqrt{s}}\frac{d
    k_t}{k_t^{1+2\epsilon}}\int_{\ln \frac{k_t}{\sqrt{s}}}^{-\cosh^{-1}\frac{\sqrt{s}}{2k_t}}d \eta \notag\\
&+ \int_{0}^{\sqrt{s}}\frac{d
    k_t}{k_t^{1+2\epsilon}}\int_{\cosh^{-1}\frac{\sqrt{s}}{2k_t}}^{\ln
 \frac{\sqrt{s}}{ k_t}}d \eta + \int_{\frac{\sqrt{s}}{2}}^{\sqrt{s}}\frac{d
    k_t}{k_t^{1+2\epsilon}}\int_{-\cosh^{-1}\frac{\sqrt{s}}{2k_t}}^{\cosh^{-1}\frac{\sqrt{s}}{2k_t}}
  d \eta \bigg] S^{\rm soft}_3(v) \Theta^{\rm soft}_{\rm out}(k)\,.
\end{align}
For the first term we can now switch to the same $x$, $y$ variables
used for Eq.~\eqref{eq:Sigmav-3jet}
\begin{equation}
\int_{0}^{\frac{\sqrt{s}}{2}}\frac{d
    k_t}{k_t^{1+2\epsilon}}\int_{-\cosh^{-1}\frac{\sqrt{s}}{2k_t}}^{\cosh^{-1}\frac{\sqrt{s}}{2k_t}}d
                                     \eta = 4^{\epsilon} \,s^{-\epsilon}\int_0^1d
                                      x \,x^{-1-2
                                      \epsilon}\int_{-1}^1 d y\, (1-y)^{-1-\epsilon} (1+y)^{-1-\epsilon}\,,
\end{equation}
while the integrals in the second line of
Eq.~\eqref{eq:Sigmav-3jet-soft-recast} are now finite and one can take
the limit $\epsilon\to 0$ prior to integration. We can now take the
difference between Eq.~\eqref{eq:Sigmav-3jet}
and~\eqref{eq:Sigmav-3jet-soft-recast} and obtain
\begin{align}
  \label{eq:Sigmav-3jet-difference}
  \Sigma^{{\rm real}}(v)&-\Sigma^{{\rm real}}_{\rm soft}(v) =
  2C_F\frac{\alpha_s}{2\pi}\left(\frac{\mu^2}{s}\right)^{\epsilon}\frac{e^{\gamma_E\epsilon}}{\Gamma(1-\epsilon)}\int_0^1d
                                      x \,x^{-1-2
                                      \epsilon}\int_{-1}^1 d y\,  (1-y)^{-1-\epsilon} (1+y)^{-1-\epsilon}\notag\\
&\!\!\!\!\!\times\bigg[\frac{8-\left(\epsilon x^2 (2-x (1-y))^2+(2-x) x \left((x-2) x (1-y)^2-4 y+8\right)\right)}{(1-x)^{2
                                      \epsilon}(2-x
   (1-y))^{2 -2\epsilon} }\Theta_{\rm out}(k) S_3(v)\notag\\
&\!\!\!\!\!-2^{1+2\epsilon}\Theta^{\rm soft}_{\rm out}(k)
  S^{\rm soft}_3(v)\bigg]\notag\\
& \!\!\!\!\!- 4C_F\frac{\alpha_s}{2\pi}\bigg[\int_{0}^{\sqrt{s}}\frac{d
    k_t}{k_t}\int_{\ln \frac{k_t}{\sqrt{s}}}^{-\cosh^{-1}\frac{\sqrt{s}}{2k_t}}d \eta + \int_{0}^{\sqrt{s}}\frac{d
    k_t}{k_t}\int_{\cosh^{-1}\frac{\sqrt{s}}{2k_t}}^{\ln\frac{\sqrt{s}}{ k_t}
 }d \eta \notag\\
&\qquad \,\,\,\,\,\,+ \int_{\frac{\sqrt{s}}{2}}^{\sqrt{s}}\frac{d
    k_t}{k_t}\int_{-\cosh^{-1}\frac{\sqrt{s}}{2k_t}}^{\cosh^{-1}\frac{\sqrt{s}}{2k_t}}d
  \eta \bigg] S_3(v) \Theta^{\rm soft}_{\rm out}(k)\,,
\end{align}
where we took the $\epsilon\to 0$ limit of the second line of
Eq.~\eqref{eq:Sigmav-3jet-soft-recast}.
We expand the integrand in the first three lines in a Laurent series
in $\epsilon$, obtaining
\begin{align}
&  \Sigma^{{\rm real}}(v)-\Sigma^{{\rm real}}_{\rm soft}(v) =
  C_F\frac{\alpha_s}{2\pi}\left(\frac{\mu^2}{s}\right)^{\epsilon}\frac{e^{\gamma_E\epsilon}}{\Gamma(1-\epsilon)}\int_0^1\frac{d x}{x} \,\int_{-1}^1 d
                           y\,\notag\\
&\times \bigg\{-\frac{\delta(1-y)+\delta(1+y)}{\epsilon}\left[(1+(1-x)^2) \Theta_{\rm out}(k) S_3(v) - 2
                           \Theta^{\rm soft}_{\rm out}(k) S^{\rm soft}_3(v) \right]\notag\\
&+\left[\frac{1}{(1-y)_+}+\frac{1}{(1+y)_+}\right]\bigg[\frac{(x-2)
  x \left((x-2) x (1-y)^2-4 y+8\right)+8}{(2-(1-y)x)^2}\Theta_{\rm
  out}(k) S_3(v) \notag\\
& - 2  \Theta^{\rm soft}_{\rm out}(k) S^{\rm soft}_3(v)
  \bigg]\notag\\
& + \delta(1-y)\big[\left(x^2 + 2 (1+(1-x)^2) \ln \left(x (1-x)\right)\right) \Theta_{\rm out}(k) S_3(v) - 4\ln(x) \Theta^{\rm
  soft}_{\rm out}(k) S^{\rm soft}_3(v) \big]\notag\\
& + \delta(1+y)\left[\left(x^2+2 (1+(1-x)^2)  \ln (x)\right) \Theta_{\rm out}(k) S_3(v) - 4\ln(x) \Theta^{\rm
  soft}_{\rm out}(k) S^{\rm soft}_3(v) \right]\bigg\}\notag\\
& - 4C_F\frac{\alpha_s}{2\pi}\bigg[\int_{0}^{\sqrt{s}}\frac{d
    k_t}{k_t}\int_{\ln \frac{k_t}{\sqrt{s}}}^{-\cosh^{-1}\frac{\sqrt{s}}{2k_t}}d \eta + \int_{0}^{\sqrt{s}}\frac{d
    k_t}{k_t}\int_{\cosh^{-1}\frac{\sqrt{s}}{2k_t}}^{\ln\frac{\sqrt{s}}{ k_t}
 }d \eta + \int_{\frac{\sqrt{s}}{2}}^{\sqrt{s}}\frac{d
    k_t}{k_t}\int_{-\cosh^{-1}\frac{\sqrt{s}}{2k_t}}^{\cosh^{-1}\frac{\sqrt{s}}{2k_t}} d \eta \bigg]\notag\\
&\times S_3(v) \Theta^{\rm soft}_{\rm out}(k)\,.
\end{align}
To proceed, we observe that the delta functions $\delta(1\pm y)$ act
as follows
\begin{align}
  \delta(1\pm y)\Theta_{\rm out}(k) S_3(v) =\delta(1\pm y)\Theta^{\rm soft}_{\rm out}(k) S^{\rm soft}_3(v) =\delta(1\pm y) S_2(v)\,,
\end{align}
since the gluon is projected onto the collinear limits. Therefore, we
can evaluate the integrals in all terms containing $\delta(1\pm y)$
and get
\begin{align}
  \label{eq:Sigmav-3jet-difference-expanded}
&  \Sigma^{{\rm real}}(v)-\Sigma^{{\rm real}}_{\rm soft}(v) =
  C_F\frac{\alpha_s}{2\pi}\left(\frac{\mu^2}{s}\right)^{\epsilon}\frac{e^{\gamma_E\epsilon}}{\Gamma(1-\epsilon)}\bigg\{\left[\frac{3}{\epsilon}+\left(\frac{21}{2}-\frac{2}{3}\pi^2\right)\right]\,S_2(v)\notag\\
& + \int_0^1\frac{d x}{x} \,\int_{-1}^1 d
                           y\,\left[\frac{1}{(1-y)_+}+\frac{1}{(1+y)_+}\right]\notag\\
&\times\bigg[\frac{(x-2)
  x \left((x-2) x (1-y)^2-4 y+8\right)+8}{(2-(1-y)x)^2}\Theta_{\rm
  out}(k) S_3(v) - 2  \Theta^{\rm soft}_{\rm out}(k) S^{\rm soft}_3(v)
  \bigg]\bigg\}\notag\\
& - 4C_F\frac{\alpha_s}{2\pi}\bigg[\int_{0}^{\sqrt{s}}\frac{d
    k_t}{k_t}\int_{\ln \frac{k_t}{\sqrt{s}}}^{-\cosh^{-1}\frac{\sqrt{s}}{2k_t}}d \eta + \int_{0}^{\sqrt{s}}\frac{d
    k_t}{k_t}\int_{\cosh^{-1}\frac{\sqrt{s}}{2k_t}}^{\ln\frac{\sqrt{s}}{ k_t}
 }d \eta + \int_{\frac{\sqrt{s}}{2}}^{\sqrt{s}}\frac{d
    k_t}{k_t}\int_{-\cosh^{-1}\frac{\sqrt{s}}{2k_t}}^{\cosh^{-1}\frac{\sqrt{s}}{2k_t}} d \eta \bigg]\notag\\
&\times S_3(v) \Theta^{\rm soft}_{\rm out}(k)\,.
\end{align}

The hard factors ${\cal H}_2$ and ${\cal H}_3$ can be directly
extracted from the above computation according to the definition given
in Eq.~\eqref{eq:convolution}, and they read
\begin{align}
\label{eq:finalH2}
{\cal H}_2 = \left[1+C_F\frac{\alpha_s}{2\pi}\left(\frac{5}{2}+\frac{\pi^2}{3}\right)\right]\delta^{(2)}(\Omega_1-\Omega_q)
  \delta^{(2)}(\Omega_2-\Omega_{\bar q}) \,,
\end{align}
and
\begin{align}
\label{eq:finalH3}
{\cal H}_3 &= C_F\frac{\alpha_s}{2\pi}\bigg\{\int_0^1\frac{d x}{x} \,\int_{-1}^1 d
                           y\,\left[\frac{1}{(1-y)_+}+\frac{1}{(1+y)_+}\right]\notag\\
&\times\bigg[\frac{(x-2)
  x \left((x-2) x (1-y)^2-4 y+8\right)+8}{(2-(1-y)x)^2} - 2
  \mathbb{P}_{\rm soft}
  \bigg] \notag\\
&- 4\bigg[\int_{0}^{\sqrt{s}}\frac{d
    k_t}{k_t}\int_{\ln \frac{k_t}{\sqrt{s}}}^{-\cosh^{-1}\frac{\sqrt{s}}{2k_t}}d \eta + \int_{0}^{\sqrt{s}}\frac{d
    k_t}{k_t}\int_{\cosh^{-1}\frac{\sqrt{s}}{2k_t}}^{\ln\frac{\sqrt{s}}{ k_t}}d \eta + \int_{\frac{\sqrt{s}}{2}}^{\sqrt{s}}\frac{d
    k_t}{k_t}\int_{-\cosh^{-1}\frac{\sqrt{s}}{2k_t}}^{\cosh^{-1}\frac{\sqrt{s}}{2k_t}} d \eta \bigg]\mathbb{P}_{\rm soft}
  \bigg\}\notag\\
&\times\, \Theta_{\rm out}(k) \,\delta^{(2)}(\Omega_1-\Omega_q)
  \delta^{(2)}(\Omega_2-\Omega_{\bar q}) \delta^{(2)}(\Omega_3-\Omega_g)\,,
\end{align}
where we introduced the projector $\mathbb{P}_{\rm soft}$ which maps
the gluon momentum $k$ into its soft limit, such that
\begin{align}
\mathbb{P}_{\rm soft}\Theta_{\rm out}(k)S_3(v)=\Theta^{\rm soft}_{\rm out}(k)
  S^{\rm soft}_3(v)\,.
\end{align}
The plus distributions act on the soft factors providing the
counterterms necessary to make the integration finite and suitable for
a numerical evaluation.  They are to be evaluated only within the
convolution integral of Eq.~\eqref{eq:convolution} as they also act on
the soft factor $S_3$.
The expressions for ${\cal H}_2$ and ${\cal H}_3$ depend on the
specific phase-space parametrisation adopted for the three-parton
final state as constant terms could be reshuffled between the two
contributions, and only the physical combination of the two is
invariant. 

As a check, we can set $\Theta_{\rm out}(k)=1$ in
eq.~\eqref{eq:finalH3} and integrate $\mathcal{H}_2$ and
$\mathcal{H}_3$ over all solid angles to obtain
\begin{align}
\int d^{2}\Omega_1\,d^{2}\Omega_2\,
{\cal H}_2+\int d^{2}\Omega_1\,d^{2}\Omega_2\,d^{2}\Omega_3\,\left.{\cal H}_3\right|_{\Theta_{\rm out}(k)=1} =
  1+\frac{3}{2} C_F\frac{\alpha_s}{2\pi}\,,
\end{align}
which reproduces the well known total cross section for $e^+e^-\to$
hadrons at NLO normalised to the Born result.
We finish this section by computing the first-order
observable-independent hard contribution to $\Sigma(v)$, defined by
\begin{equation}
  \label{eq:H1-defined}
\int d^{2}\Omega_1\,d^{2}\Omega_2\,
{\cal H}_2+\int d^{2}\Omega_1\,d^{2}\Omega_2\,d^{2}\Omega_3\,{\cal H}_3 \coloneqq
  1+ C_F\frac{\alpha_s}{2\pi} H_1(c)\,,
\end{equation}
where, to compute $\mathcal{H}_3$, we need to use the explicit
expression for $\Theta_{\rm out}(k)$, as follows
\begin{equation}
  \label{eq:Theta-out}
  \Theta_{\rm out}(k)=\Theta\left(\mathrm{med}\left[1-\cos\theta_{q\bar q},1-\cos\theta_{qg},1-\cos\theta_{\bar qg}\right]-(1+c)\right)\,.
\end{equation}
As pointed out already in Ref.~\cite{Becher:2016mmh}, the result depends only on the quantity
\begin{equation}
  \label{eq:delta-c}
  \delta \coloneqq \tan\frac{\theta_{\rm jet}}{2} = \sqrt\frac{1-c}{1+c}=e^{-\Delta\eta/2}\,.
\end{equation}
In particular, it depends only on $\delta^2$. For $c\ge 1/2$, we
obtain
\begin{equation}
  \label{eq:H3-largec}
  \begin{split}
  H_1(c)=-1+6\ln 2+\frac{\pi^2}{3}-3\ln(\delta^2)+\ln^2(\delta^2)-6\delta^2+\left(\frac{9}{2}-6\ln 2\right)\delta^4+4 \mathrm{Li}_2\left(\delta^2\right)\,.
  \end{split}
\end{equation}
Eq.~\eqref{eq:H1-defined} only accounts for the hard contribution,
while there will be an additional constant term at
${\cal O}(\alpha_s)$ coming from $S_2$.
In the case of the energy distribution, plugging
Eq.~\eqref{eq:H1-defined} in eq.~(\ref{eq:master}), and expanding at
first order with $v(k)=2 \omega$ leads to the same result as
Ref.~\cite{Becher:2016mmh}.
We stress that this comparison must necessarily be carried out at the
level of the physical quantity $H_1(c)$ and not individually for
${\cal H}_2$ and ${\cal H}_3$ which, as mentioned above, depend on the
scheme used to cancel the collinear singularities between the two as
well as on the definition of the functions $S_{2}$ and $S_3$.
The region $c<1/2$ is not considered in Ref.~\cite{Becher:2016mmh}. We obtain 
\begin{equation}
  \label{eq:H3-smallc}
  \begin{split}    
  H_1(c)& =-6+12 \delta^2-\left(\frac{9}{2}-6\ln\frac{2 \delta^2}{1+\delta^2}\right)\delta^4+2\ln^2 2 -6 \ln 2 - 9 \ln(\delta^2)-\ln^2(\delta^2)\\ & +4 \ln(\delta^2)\ln(1-\delta^2)+6 \ln(1+\delta^2)-4\ln 2 \ln(1+\delta^2)-4 \ln\delta^2\ln(1+\delta^2) \\ &+2 \ln^2(1+\delta^2) +4\mathrm{Li}_2\left(\frac{1-\delta^2}{2}\right)+4 \mathrm{Li}_2 \left(\frac{1-\delta^2}{1+\delta^2}\right)\,.
  \end{split}
\end{equation}

\section{NLL corrections up to ${\cal O}(\alpha_s^2)$ and fixed order
  tests}
\label{eq:event2}
In order to test our result Eq.~\eqref{eq:master}, we perform a fixed
order expansion of $\Sigma(v)$ up to ${\cal O}(\alpha_s^2)$. Here we
set the evolution scale as $Q=\sqrt{s}$. Variations of $Q$ around
$\sqrt{s}$ would correspond to standard variations of the resummation
scale (used to assess the corresponding theoretical uncertainty), that
we do not consider here in the context of a fixed-order expansion. We
first start using the explicit expressions for $S_2(v)$ and $S_3(v)$
in the large-$N_c$ limit. Once the large-$N_c$ result is established,
we upgrade it to include finite-$N_c$ corrections.
Working at NLL accuracy, we obtain
\begin{equation}
  \label{eq:master-explicit}
  \Sigma(v) ={\cal H}_2  \otimes \int\frac{d\nu}{2\pi i \nu}e^{\nu v}
  G_{12}[Q;u] + {\cal H}_3 \otimes \int\frac{d\nu}{2\pi i \nu} e^{\nu v} G_{13}[Q;u] G_{32}[Q;u]\,. 
\end{equation}
where $p_1,p_2,p_3$ are the momenta of the final-state quark,
antiquark and gluon. In general, $G_{ij}[Q;u]$ ($ij=12,13,32$) can be
written as an expansion in powers of $\abar=\abar(Q)$, as follows
\begin{equation}
  \label{eq:Gij-expanded}
  G_{ij}[Q;u]=1+\abar\, G_{ij}^{(1)}[Q,u]+\abar^2 \,G_{ij}^{(2)}[Q,u]+\dots \,,
\end{equation}
where
\begin{equation}
  \label{eq:Gij-exp-terms}
  \begin{split}
    G_{ij}^{(1)}[Q&,u]  = \int [dk] w^{(0)}_{ij}(k) \left[u(k)-1\right]\Theta(Q-k_t)\,,\\
    G_{ij}^{(2)}[Q&,u]  = \frac{1}{2}\left(G_{ij}^{(1)}[Q,u]\right)^2+ \int [dk] w^{(0)}_{ij}(k)[u(k)-1]\Theta(Q-k_t)\left(\bar K^{(1)}-2\bar\beta_0 \ln\frac{k_{t}}{Q} \right) \\
    &+\int [dk_a]\int [dk_b] \left[\bar
      w^{(gg)}_{ij}(k_a,k_b)+\bar
      w^{(gg)}_{ij}(k_b,k_a))\right]\Theta(Q-k_{t(ab)})\,\Theta(k_{ta}-k_{tb}^\prime) \times \\ & \qquad\qquad\times
 \left[u(k_a)u(k_b)-u(k_{(ab)})\right]\,.
  \end{split}
\end{equation}
Similarly, also ${\cal H}_2$ and ${\cal H}_3$ can be expanded in
powers of $\alpha_s$, with the convention of
eq.~(\ref{eq:conventions}). Therefore, using the explicit expression
for $\abar=N_c \alpha_s/\pi$, and keeping all terms up to order
$\alpha_s^2$, we obtain
\begin{align}
  \label{eq:master-expanded}
  \Sigma(v)&=1+\frac{\alpha_s}{2\pi} \left(2 N_c\int\frac{d\nu}{2\pi i \nu}e^{\nu v}
     G^{(1)}_{12}[Q;u] +{\cal H}^{(1)}_2 + {\cal H}^{(1)}_3 \otimes
    \mathbbm{1} \right) \notag\\
& \qquad\qquad + \left(\frac{\alpha_s}{2\pi}\right)^2 \int\frac{d\nu}{2\pi i
  \nu}e^{\nu v} \bigg[4 N_c^2 G^{(2)}_{12}[Q;u]+2 N_c \,{\cal
  H}_2^{(1)} G_{12}^{(1)}[Q;u]\\
&\qquad \qquad\qquad +2 N_c\, {\cal H}^{(1)}_3 \otimes
  \left(G_{13}^{(1)}[Q;u]+G_{32}^{(1)}[Q;u]\right)\bigg] + {\cal O}(\alpha_s^3)\,.\notag
\end{align}
We now compute all inverse Laplace transforms by observing that they
affect only the sources, and not the matrix element squared or the
phase space. For $u(k)$ as in eq.~(\ref{eq:source}), we obtain
\begin{align}
  \label{eq:inverse-Laplace}
    \int\frac{d\nu}{2\pi i \nu}e^{\nu v} u(k)& =\Theta_{\rm out}(k) + \Theta_{\rm in}(k) \Theta(v-v(k))\,. \\
    \int\frac{d\nu}{2\pi i \nu}e^{\nu v} u(k_a)u(k_b)& =\Theta_{\rm out}(k_a)\Theta_{\rm out}(k_b) + 
    \Theta_{\rm in}(k_a)\Theta_{\rm in}(k_{b})\Theta(v-v(k_{a})-v(k_{b}))\notag\\ & +
    \Theta_{\rm in}(k_a)\Theta_{\rm out}(k_{b}) \Theta(v-v(k_{a}))
    + \Theta_{\rm out}(k_a)\Theta_{\rm in}(k_{b}) \Theta(v-v(k_{b}))\,.\notag
\end{align}
We use the above information to compute separately each term that
depends on the sources in eq.~\eqref{eq:master-expanded}. Introducing
$L\coloneqq \ln(Q/v)=\{\ln Q/E, \ln Q/E_t\}$, we obtain:
\begin{align}
  \label{eq:Gij-exp-invLaplace-1}
  \int\frac{d\nu}{2\pi i \nu}e^{\nu v} G^{(1)}_{12}[Q;u]  &= -\int [dk] w_{12}^{(0)}(k)
                                                            \Theta_{\rm in}(k) \Theta(v(k)-v)\Theta(Q-k_t)\,,%=-\Delta \eta \,L\,,
\end{align}
and
\begin{align}
    \label{eq:Gij-exp-invLaplace-2}
\int\frac{d\nu}{2\pi i \nu}e^{\nu v} G^{(2)}_{12}[Q;u]  &= -\int [dk] w^{(0)}_{12}(k)\Theta_{\rm in}(k) \Theta(v(k)-v)\Theta(Q-k_t)\left(\bar K^{(1)} -2\bar\beta_0 \ln\frac{k_{t}}{Q}\right)\notag\\
  & +  \int [dk_a] [dk_b] w_{12}^{(0)}(k_a) w_{12}^{(0)}(k_b)
  \Theta_{\rm in}(k_a) \Theta_{\rm in}(k_b)\Theta(Q-k_{ta})\Theta(k_{ta}-k_{tb})\notag\\
&\times\left[ \Theta(v(k_{b})-v)+\Theta(v-v(k_{a})-v(k_{b}))-\Theta(v-v(k_{a}))\right]\notag\\
  & +\int [dk_a]\int [dk_b] \left[\bar w^{(gg)}_{12}(k_a,k_b)+\bar
    w^{(gg)}_{12}(k_b,k_a)\right] \Theta(Q-k_{t(ab)})\Theta(k_{ta}-k_{tb}^\prime)\notag\\
&\times \left[\Theta_{\rm out}(k_a)\Theta_{\rm out}(k_b) + 
    \Theta_{\rm in}(k_a)\Theta_{\rm in}(k_{b})\Theta(v-v(k_{a})-v(k_{b})) \right.\notag\\ & \left.+
    \Theta_{\rm in}(k_a)\Theta_{\rm out}(k_{b}) \Theta(v-v(k_{a}))
    + \Theta_{\rm out}(k_a)\Theta_{\rm in}(k_{b}) \Theta(v-v(k_{b}))\right.\notag\\
&  \left.  -\Theta_{\rm out}(k_{(ab)})-\Theta_{\rm in}(k_{(ab)})\Theta(v-v(k_{(ab)}))\right]\,.
\end{align}
The term containing $ w_{12}^{(0)}(k_a) w_{12}^{(0)}(k_b)$ corresponds
to independent emissions. There, when both emissions are inside
the slice, the constraint
$\Theta(v-v(k_{a})-v(k_{b}))-\Theta(v-v(k_{a}))$ gives a contribution
of order $\abar^2$, without any logarithmic enhancement. Therefore, we
can neglect that term and obtain, up to NNLL corrections,
\begin{equation}
  \label{eq:indep-ngls}
  \begin{split}
 & \int [dk_a] [dk_b] w_{12}^{(0)}(k_a) w_{12}^{(0)}(k_b) \Theta_{\rm in}(k_a) \Theta_{\rm in}(k_b)\Theta(Q-k_{ta})\Theta(k_{ta}-k_{tb})\Theta(v(k_{b})-v) \\& \qquad= \frac 12 \left(\int [dk] w_{12}^{(0)}(k)
  \Theta_{\rm in}(k) \Theta(v(k)-v)\Theta(Q-k_t)\right)^2\,.% = \frac 12 (\Delta \eta \,L)^2\,.
  \end{split}
\end{equation}
The term containing
$\bar w^{(gg)}_{12}(k_a,k_b)+\bar w^{(gg)}_{12}(k_b,k_a)$ corresponds
to correlated emissions. This is the term that gives rise to leading
non-global logarithms at this perturbative order. The observable
constraints can be rearranged as follows
\begin{multline}
  \label{eq:ngl-obs-constraints}
  -\Theta_{\rm in}(k_b)\Theta_{\rm out}(k_a)\Theta(v(k_{b})-v)\\
  -\Theta_{\rm in}(k_b)\left[\Theta_{\rm in}(k_a)\Theta(v(k_{a})+v(k_{b})-v)-\Theta_{\rm in}(k_{(ab)}) \Theta(v(k_{(ab)})-v)\right]\\
  -\Theta_{\rm out}(k_b) \left[\Theta_{\rm in}(k_a)\Theta(v(k_{a})-v)-\Theta_{\rm in}(k_{(ab)}) \Theta(v(k_{(ab)})-v)\right]\,.
\end{multline}
Each line in the above equation gives rise to a finite contribution,
without soft or collinear divergences. The term in the first line
gives rise to leading non-global logarithms. The term in the second
line can be further simplified by observing that, when both
$k_a$ and $k_b$ are inside the slice, the parent $k_{(ab)}$ is forced
by kinematics to be inside the slice as well. Therefore, the only
non-zero contribution to the second line of
eq.~\eqref{eq:ngl-obs-constraints} is given by
\begin{multline}
  \label{eq:ngl-obs-bothin}
  -\Theta_{\rm in}(k_{(ab)}) \Theta_{\rm in}(k_b)\left[\Theta_{\rm in}(k_a)\Theta(v(k_{a})+v(k_{b})-v)-\Theta(v(k_{(ab)})-v)\right] \\= \Theta_{\rm in}(k_{(ab)}) \Theta_{\rm in}(k_b)\Theta_{\rm out}(k_a)\Theta(v(k_{(ab)})-v)\\
  -\Theta_{\rm in}(k_{(ab)}) \Theta_{\rm in}(k_b)\Theta_{\rm in}(k_a)\left[\Theta(v(k_{a})+v(k_{b})-v)-\Theta(v(k_{(ab)})-v)\right]\,.
\end{multline}  
The last line of the above equation corresponds to a collinear and
infrared finite contribution, where the phase-space of the parent
gluon is integrated over a region where its rapidity is finite and its
transverse momentum bounded from above and from below by two
quantities of order $v$, hence giving clearly a finite contribution of
order $\alpha_s^2$. Similarly, to further simplify
eq.~\eqref{eq:ngl-obs-constraints}, we keep track of whether the
parent $k_{(ab)}$ is inside or outside the slice. This gives
\begin{multline}
  \label{eq:ngl-obs-simp}
  -\Theta_{\rm out}(k_{(ab)})\left[\Theta_{\rm out}(k_a)\Theta_{\rm in}(k_b)\Theta(v(k_{b})-v)+\Theta_{\rm in}(k_a)\Theta_{\rm out}(k_b)\Theta(v(k_{a})-v)\right]\\
  +\Theta_{\rm in}(k_{(ab)})\Theta_{\rm out}(k_a)\Theta_{\rm out}(k_b)\Theta(v(k_{(ab)})-v)\\
  -\Theta_{\rm in}(k_{(ab)})\Theta_{\rm out}(k_a)\Theta_{\rm in}(k_b)\left[\Theta(v(k_{b})-v)-\Theta(v(k_{(ab)})-v)\right]\\
    -\Theta_{\rm in}(k_{(ab)})\Theta_{\rm in}(k_a)\Theta_{\rm out}(k_b)\left[\Theta(v(k_{a})-v)-\Theta(v(k_{(ab)})-v)\right]\,.
  \end{multline}
  Again, the contributions in the last two lines of the above equation
  corresponds to phase space regions in which the rapidity of the
  parent gluon is bounded (it is in fact inside the gap, and its
  transverse momentum bounded by two limits of order $v$). Such
  regions give contributions of order $\alpha_s^2$ (with no
  logarithmic enhancement) and can therefore be neglected. Therefore,
  the only regions of phase space where correlated soft gluon emission
  can give rise to logarithmically enhanced contributions are those
  where the parent is outside the gap and only one of its offspring is
  inside the slice and vetoed, and another where the parent is inside
  the gap and vetoed and both $k_a$ and $k_b$ are outside. This gives
  the final contribution due to correlated emission:
  \begin{align}
    \label{eq:correl-ngls}
    \int [dk_a]&\int [dk_b] \left[\bar w^{(gg)}_{12}(k_a,k_b)+\bar
      w^{( gg)}_{12}(k_b,k_a)\right] \Theta(Q-k_{t(ab)})\Theta(k_{ta}-k_{tb}^\prime) \\ 
&\times
\left\{
  \Theta_{\rm in}(k_{(ab)})\Theta_{\rm out}(k_a)\Theta_{\rm out}(k_b)\Theta(v(k_{(ab)})-v)\right.
\notag\\ & \left. 
-\Theta_{\rm out}(k_{(ab)})\left[\Theta_{\rm in}(k_a)\Theta_{\rm out}(k_b)\Theta(v(k_{a})-v)+\Theta_{\rm out}(k_a)\Theta_{\rm in}(k_b)\Theta(v(k_{b})-v)\right]
\right\}\notag
\end{align}
Altogether we obtain, up to NNLL corrections
\begin{align}
  \label{eq:G12-2-simplified}
  \int\frac{d\nu}{2\pi i \nu}e^{\nu v} &G^{(2)}_{12}[Q;u]  \simeq-\int
  [dk_a] w^{(0)}_{12}(k)\Theta_{\rm in}(k)
  \Theta(v(k)-v)\Theta(k_{t}-Q)\left(\bar K^{(1)} -2\bar\beta_0
    \ln\frac{k_{t}}{Q}\right)\notag\\ 
& +\frac 12 \left(\int [dk] w_{12}^{(0)}(k)
    \Theta_{\rm in}(k) \Theta(v(k)-v)\Theta(Q-k_t)\right)^2\notag\\
  & +    \int [dk_a]\int [dk_b] \left[\bar w^{(gg)}_{12}(k_a,k_b)+\bar
      w^{( gg)}_{12}(k_b,k_a)\right] \Theta(Q-k_{t(ab)})\Theta(k_{ta}-k_{tb}^\prime) \notag\\ 
&\times
\left\{
  \Theta_{\rm in}(k_{(ab)})\Theta_{\rm out}(k_a)\Theta_{\rm out}(k_b)\Theta(v(k_{(ab)})-v)\right.\\ & \left. 
-\Theta_{\rm out}(k_{(ab)})\left[\Theta_{\rm in}(k_a)\Theta_{\rm out}(k_b)\Theta(v(k_{a})-v)+\Theta_{\rm out}(k_a)\Theta_{\rm in}(k_b)\Theta(v(k_{b})-v)\right]
\right\}\,.\notag
\end{align}
The last contributions we need to compute at order $\alpha_s^2$ are those involving convolutions of the hard factors $\mathcal{H}_2$ and $\mathcal{H}_3$ with $G_{ij}^{(1)}[Q,u]$. This gives
\begin{align}
  \label{eq:H2-H3-G}
  \int&\frac{d\nu}{2\pi i \nu}e^{\nu v}\left[{\cal H}_2^{(1)}
                                        G_{12}^{(1)}[Q;u]+{\cal
                                        H}^{(1)}_3 \otimes
                                        \left(G_{13}^{(1)}[Q;u]+G_{32}^{(1)}[Q;u]\right)\right]\\
&=-\int [dk] 
  \Theta_{\rm in}(k) \Theta(v(k)-v)\Theta(Q-k_t)\left[ {\cal H}_2^{(1)} w_{12}^{(0)}(k)+
  {\cal H}^{(1)}_3\otimes \left(w_{13}^{(0)}(k)+w_{32}^{(0)}(k)\right)\right]\,.\notag
\end{align}
To summarise, the expansion of $\Sigma(v)$ up to order $\alpha_s^2$
can be obtained by inserting
Eqs.~(\ref{eq:Gij-exp-invLaplace-1}),~(\ref{eq:G12-2-simplified})
and~(\ref{eq:H2-H3-G}) into Eq.~(\ref{eq:master-expanded}). 

\begin{figure}[htbp]
  \centering
  \includegraphics[width=.8\textwidth]{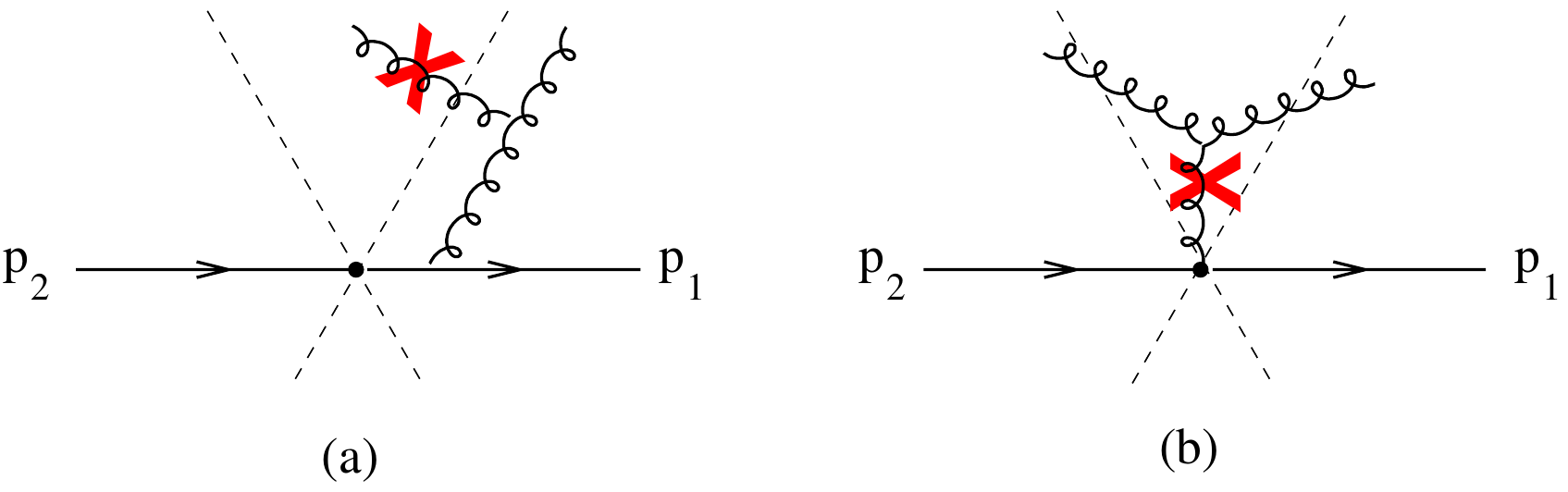}
  \caption{The two representative contributions corresponding to two
    soft correlated emissions. The red cross corresponds to the
    constraint $-\Theta(v(k)-v)$ on soft gluon $k$.}
  \label{fig:2emsn-ngls}
\end{figure}

\begin{figure}[htbp]
  \centering
  \includegraphics[width=\textwidth]{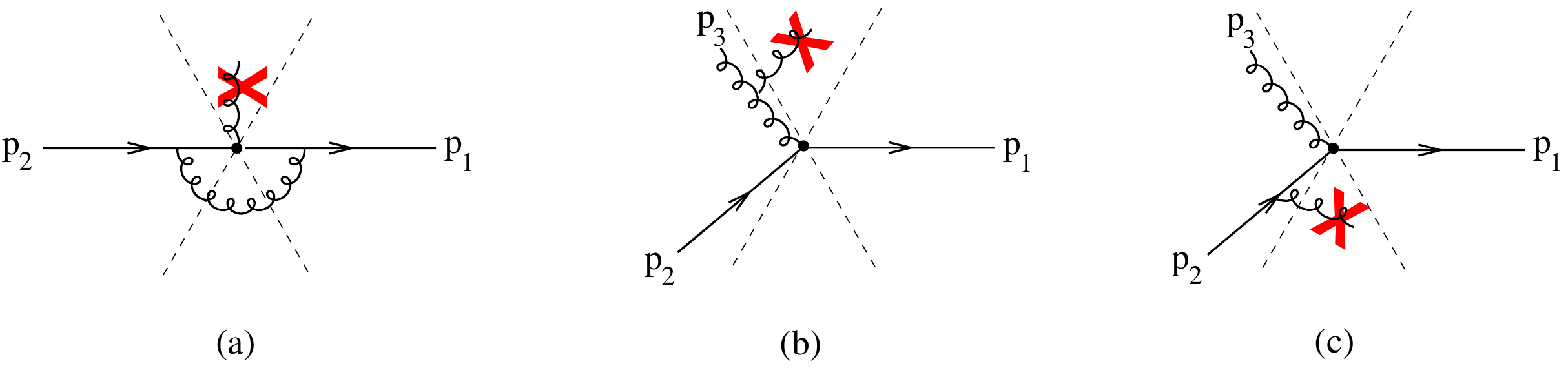}
  \caption{Representative configurations corresponding to a soft gluon
    emitted from hard $\mathcal{O}(\alpha_s)$ configurations. The red
    cross corresponds to the constraint $-\Theta(v(k)-v)$ on soft
    gluon $k$. }
  \label{fig:3jets-ngls}
\end{figure}

\paragraph{Extension to finite $N_c$.}
Since we ultimately wish to compare to an exact fixed-order
calculation, we need to upgrade our $\mathcal{O}(\alpha_s^2)$ results
to finite $N_c$.  Since we know the full expression of the matrix
element squared for the emission of two soft partons, as well as of
one-loop virtual corrections to the soft-gluon current, at this fixed
order (but not at higher orders) we can upgrade
eq.~(\ref{eq:master-expanded}) to finite $N_c$ by performing the
following modifications:
\begin{itemize}
\item Add the $n_f$-contribution to $\bar \beta_0$ and $\bar K^{(1)}$,
  as follows
  \begin{equation}
    \label{eq:beta0-K-finiteNc}
    \begin{split}
      \bar\beta_0  \to \frac{\beta_0}{N_c}\pi\,,\qquad
      \bar K^{(1)} \to \frac{K^{(1)}}{2 \,N_c}\,,
    \end{split}
  \end{equation}
  where $\beta_0=(11 C_A-2\,n_f)/(12 \pi)$ and
$K^{(1)}=\left(67/18-\pi^2/6\right)C_A -5/9 \,n_f$.
\item Add the $n_f$-contribution to $\bar w_{12}^{(gg)}(k_a,k_b)$, as follows
  \begin{equation}
    \label{eq:w12-kakb-nf}
    \bar w_{12}^{(gg)}(k_a,k_b)\to \bar w_{12}^{(gg)}(k_a,k_b) + \frac{n_f}{N_c} \bar{w}_{12}^{(q\bar q)}(k_a,k_b)\,,
  \end{equation}
  where we defined $\bar{w}_{12}^{(q\bar q)}$ in $4-2\epsilon$
  dimensions as (see e.g.~\cite{GehrmannDeRidder:2005cm})
\begin{align}
\bar{w}_{12}^{(q\bar q)}(k_a,k_b) =
2\,(2\pi)^4\mu^{4\epsilon}\bigg[&\frac{1}{s_{ab}^2
  (s_{1a}+s_{1b})(s_{2a}+s_{2b})}\left(s_{12}s_{ab}-s_{1a}s_{2b}-s_{2a}s_{1b}\right)\notag\\
&+ \frac{1}{s_{ab}^2}\left(\frac{s_{1a}s_{1b}}{(s_{1a}+s_{1b})^2}+\frac{s_{2a}s_{2b}}{(s_{2a}+s_{2b})^2}\right)\bigg]\,.
\end{align}

\item Add a colour-suppressed term to the contribution containing
  $\mathcal{H}_3^{(1)}$ in eq.~(\ref{eq:H2-H3-G}), corresponding to the
  emission of a soft gluon from the $q\bar q$ dipole, as follows
  \begin{equation}
    \label{eq:H3-finite-Nc}
    {\cal H}^{(1)}_3\otimes \left(w_{13}^{(0)}(k)+w_{32}^{(0)}(k)\right)\to
    {\cal H}^{(1)}_3\otimes \left(w_{13}^{(0)}(k)+w_{32}^{(0)}(k)-\frac{1}{N_c^2}w_{12}^{(0)}(k)\right)\,.
  \end{equation}

\item Replace each power of $N_c$ with the appropriate colour factor. 
\end{itemize}
Performing these upgrades gives
\begin{align}
  \label{eq:Sigma-exp-fullcolour}
      \Sigma(v)& \simeq 1+ \left(\frac{\alpha_s}{2\pi}\right)\left({\cal H}^{(1)}_2-4 C_F\int [dk] w_{12}^{(0)}(k)
      \Theta_{\rm in}(k) \Theta(v(k)-v)\Theta(Q-k_t)+ {\cal H}^{(1)}_3 \otimes
    \mathbbm{1}\right)\notag\\
               &-4 C_F \left(\frac{\alpha_s}{2\pi}\right)^2\int [dk] w^{(0)}_{12}(k)\Theta_{\rm in}(k) \Theta(v(k)-v)\Theta(k_{t}-Q)\left(K^{(1)} -4\pi\beta_0\ln\frac{k_{t}}{Q}\right)\notag\\
        & +8 C_F^2\left(\frac{\alpha_s}{2\pi}\right)^2\left(\int [dk] w_{12}^{(0)}(k)
        \Theta_{\rm in}(k) \Theta(v(k)-v)\Theta(Q-k_t)\right)^2\notag\\
      & -8 C_F \left(\frac{\alpha_s}{2\pi}\right)^2     \int [dk_a]\int [dk_b] \left[C_A\left(\bar w^{(gg)}_{12}(k_a,k_b)+\bar
      w^{(gg)}_{12}(k_b,k_a)\right)\right.\notag\\
&\left.\hspace{5cm} +n_f\left(\bar w^{(q\bar q)}_{12}(k_a,k_b)+\bar
      w^{(q\bar q)}_{12}(k_b,k_a)\right) \right] \\ 
&\times
 \Theta(Q-k_{t(ab)})\Theta(k_{ta}-k_{tb}^\prime)
\left\{
  \Theta_{\rm out}(k_{(ab)})\left[\Theta_{\rm in}(k_a)\Theta_{\rm out}(k_b)\Theta(v(k_{a})-v) \right.\right.
\notag\\ & \left.\left. +\Theta_{\rm out}(k_a)\Theta_{\rm in}(k_b)\Theta(v(k_{b})-v)\right]
-\Theta_{\rm in}(k_{(ab)})\Theta_{\rm out}(k_a)\Theta_{\rm out}(k_b)\Theta(v(k_{(ab)})-v)
           \right\}\notag\\
& -2\left(\frac{\alpha_s}{2\pi}\right)^2 \int [dk] 
\Theta_{\rm in}(k) \Theta(v(k)-v)\Theta(Q-k_t) \notag\\ 
& \times \left[ 2 C_F{\cal H}_2^{(1)} w_{12}^{(0)}(k) + {\cal H}^{(1)}_3\otimes \left( C_A (w_{13}^{(0)}(k)+ w_{32}^{(0)}(k))+(2 C_F-C_A)w_{12}(k)\right)\right]\,.\notag
\end{align}
The configurations contributing to the first two lines correspond to
the emission of a single gluon, together with the corresponding
virtual corrections. The third line is the contribution of the
independent emission of two gluons.  This is followed by the
contribution of two correlated soft emissions, corresponding to the
two configurations shown in Fig.~\ref{fig:2emsn-ngls} (and the
analogous contributions with a soft quark-antiquark
pair). Fig.~\ref{fig:2emsn-ngls}a represents configurations where
either gluon is in the rapidity slice, and the parent is
outside. Fig.~\ref{fig:2emsn-ngls}b represents instead the
configurations where the parent is inside the slice, and its offspring
is outside.
The contribution in the last line of
eq.~\eqref{eq:Sigma-exp-fullcolour} corresponds to hard partons
outside the slice emitting a soft gluon inside the slice (the terms
containing $\mathcal{H}_3^{(1)}$), and the corresponding virtual
corrections (the term containing $\mathcal{H}_2^{(1)}$). This
contribution is symbolically represented in  Fig.~\ref{fig:3jets-ngls}. 
In particular, Fig.~\ref{fig:3jets-ngls}a represents the term
containing $\mathcal{H}_2^{(1)}$, and Fig.~\ref{fig:3jets-ngls}b
corresponds to the part of the term containing $\mathcal{H}_3^{(1)}$
where soft gluon $k$ is emitted by the two dipoles containing hard
gluon $p_3$. Finally, Fig.~\ref{fig:3jets-ngls}c represents the part
of the term containing $\mathcal{H}_3^{(1)}$ where soft gluon $k$ is
emitted by the quark-antiquark dipole.

\begin{figure}
\begin{center}
\includegraphics[width=0.49\textwidth]{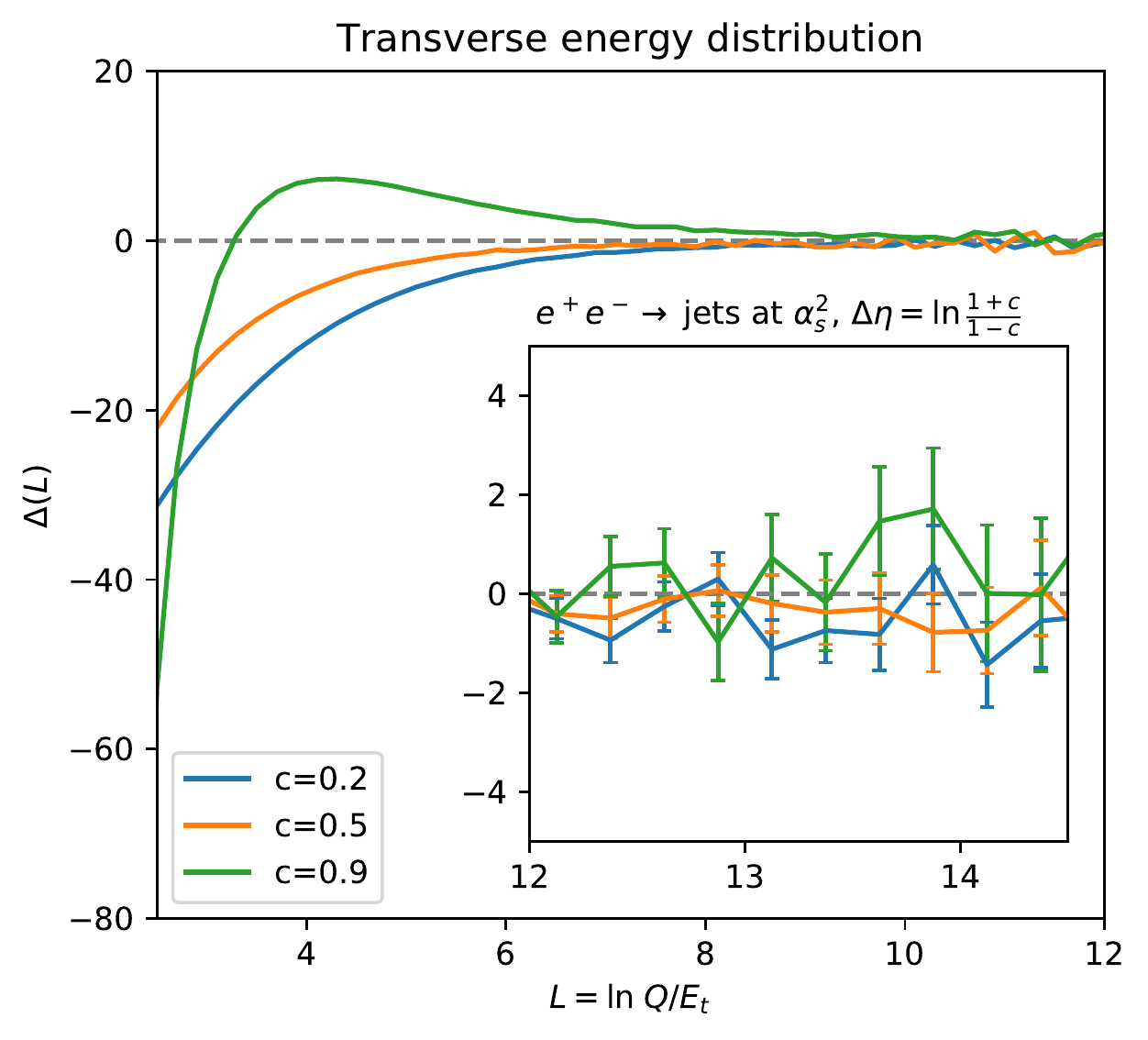}
\includegraphics[width=0.49\textwidth]{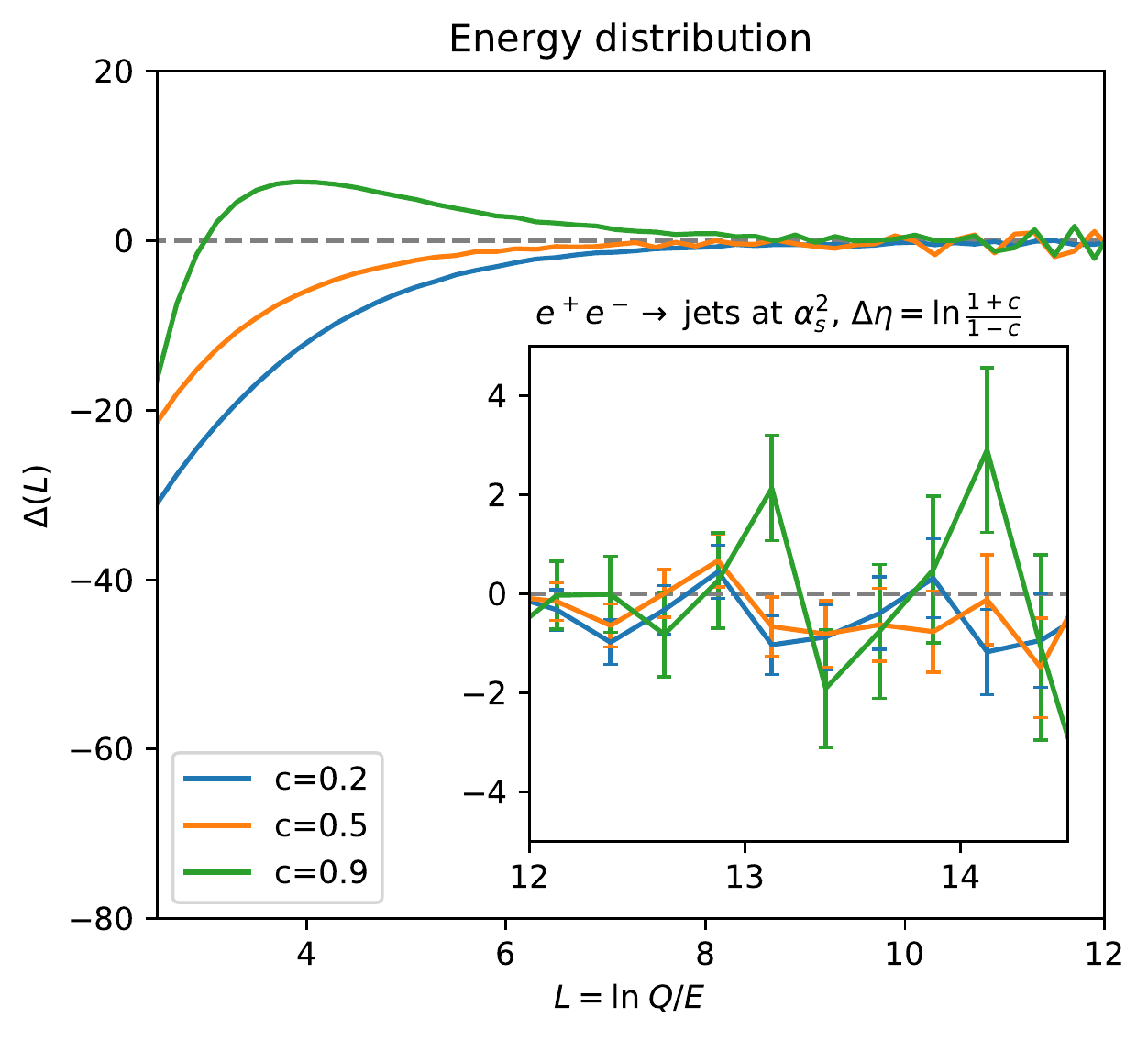}
\end{center}
\caption{\label{fig:tests} Fixed order tests against {\sc Event2} for
  transverse energy (left) and energy (right) distributions and
  different widths of the rapidity slice.}
\end{figure}

\paragraph{Comparison with \textsc{Event2}.}
We now compare the differential distribution in $v$ obtained from the
expansion in eq.~\eqref{eq:Sigma-exp-fullcolour} to the full QCD
result obtained with the program {\sc
  Event2}~\cite{Catani:1996vz}. Specifically, we consider both the
transverse energy as well as the energy distribution in the rapidity
slice of width $\Delta\eta$ and we plot the quantity
\begin{equation}
\label{eq:delta}
\Delta(L) \coloneqq \frac{1}{\sigma_0}\left(\frac{d\Sigma^{\rm NLO}}{d L}-\frac{d\Sigma^{\rm EXP.}}{d L}\right)\,, 
\end{equation}
where $d\Sigma^{\rm NLO}$ is the $\alpha_s^2/(2\pi)^2$ coefficient of
the NLO distribution extracted from {\sc Event2} and
$d\Sigma^{\rm EXP.}$ is the $\alpha_s^2/(2\pi)^2$ coefficient the
derivative of Eq.~\eqref{eq:Sigma-exp-fullcolour}.
Eq.~\eqref{eq:delta} is shown in Fig.~\ref{fig:tests} for different
values of the parameter $c=\cos\theta_{\rm jet}$.
At NLL, one expects 
\begin{equation}
\lim_{L\to \infty}\Delta(L)=0\,,
\end{equation} 
as is clearly confirmed by Fig.~\ref{fig:tests}, which validates our
predictions for the next-to-leading non-global corrections, up to
$\order{\alpha_s^2}$.
In the case of the energy distribution and $c > 1/2$, we also
compared the result of our calculation to the analytic result of
Ref.~\cite{Becher:2016mmh} and found agreement up to
${\cal O}(\alpha_s^2)$ for the values of $\delta$ adopted there.

\section{Conclusions}
\label{eq:conclusions}
In this article we have presented a formalism for the resummation of
next-to-leading non-global logarithms in QCD.  Problems sensitive to
non-global logarithms are ubiquitous in collider physics, whenever one
considers observables which are sensitive to QCD radiation only in
limited angular regions of the phase space.
As a case study, we have considered both the transverse energy and
energy distribution in a rapidity slice with respect to the thrust
axis in $e^+e^-$ annihilation.
We showed that the resummed cumulative cross section~\eqref{eq:master}
can be expressed as a sum of convolutions between hard factors
(encoding the contribution of hard radiation) and soft factors (that
resum the soft radiative corrections).
While the hard factors do not contain large logarithms, the soft
factors contain logarithmically enhanced corrections that are resummed
by a set of non-linear evolution equations in the large-$N_c$ limit.
We have computed all ingredients necessary for the resummation of the
NLL non-global corrections, namely the hard factors ${\cal H}_{2}$,
${\cal H}_{3}$ up to $\order{\alpha_s}$
(cf. Eqs.~\eqref{eq:finalH2},~\eqref{eq:finalH3}), as well as the
evolution kernel for the resummation of the soft factors $S_{2}$,
$S_{3}$ up to $\order{\alpha_s^2}$
(cf. Eqs.~\eqref{eq:LL-evolution-kt-diff-symbolic},~\eqref{eq:NLL-evolution-kt-diff}).
We used these results to carry out a ${\cal O}(\alpha_s^2)$ fixed-order calculation at finite $N_c$ that is in excellent agreement with
the full QCD prediction at ${\cal O}(\alpha_s^2)$ for asymptotically small values of the considered observable.

A natural step forward will be the all-order solution of the evolution
equations presented here and thus the resummation of NLL non-global
corrections with the corresponding study of their phenomenological
impact. We envision that the most efficient way to achieve this is by
means of Monte Carlo technology. This will be addressed in detail in
a forthcoming publication.
We finally stress that the formulation derived in this article is not
tailored to a specific observable and thus can be applied to other
infrared safe observables sensitive to soft radiation emitted away
from the large energy flow of the scattering process.
In particular, the entire process dependence is encoded in the hard
factors while the evolution of the soft factors is universal, which
will ultimately allow the resummation of next-to-leading non-global
logarithms for hadron collider observables. The precise calculation of
these corrections enters the precise description of several collider
observables involving jets. An notable example is the fraction of
gluon-fusion events in Higgs production in association with two jets
with VBF selections cuts~\cite{Hatta:2020wre}, or observables defined
by means of jet substructure
technology~\cite{Rubin:2010fc,Neill:2018yet,Lifson:2020gua}.

The application to observables sensitive to collinear radiation
(e.g. the light hemisphere mass in $e^+e^-$), on the other hand,
requires extra care since our formulation does not immediately apply
in these cases. One possible way forward would be the calculation of
the global corrections using standard resummation technology
supplemented by a non-global correction factor obtained by subtracting
the global contributions during the numerical integration of the
evolution equation derived in this article. This will require a
careful extension of the Monte Carlo method of
Ref.~\cite{Dasgupta:2001sh}, where this subtraction is carried out at
LL accuracy.
Another interesting future direction will concern the comparison of
our formalism to the formulations of
Refs.~\cite{Caron-Huot:2015bja,Becher:2016mmh}. This can be done by
explicitly computing the evolution kernels in
Eqs.~\eqref{eq:LL-evolution-kt-diff-symbolic},~\eqref{eq:NLL-evolution-kt-diff}
for a specific source corresponding to a given observable.
Moreover, from a theoretical point of view it could be interesting to
consider the inclusion of subleading-$N_c$ corrections to the
evolution kernel. Although such corrections have been observed to be
usually small in known cases~\cite{Hatta:2020wre,Hamilton:2020rcu},
their theoretical understanding must be improved in view of high
precision phenomenology.

\section*{Acknowledgments}
We are particularly grateful to Thomas Becher, Mrinal Dasgupta,
Lorenzo Magnea and Gavin Salam for insightful comments and
suggestions, and we would like to thank Keith Hamilton and Gregory
Soyez for useful discussions. We also thank Keith Hamilton for
providing an independent derivation of the decomposition of the double
soft squared current in the colour flow basis.
This work was supported by the Science Technology and Facilities
Council (STFC) under grants number ST/P000819/1, ST/T00102X/1 (AB) and
ST/T000864/1 (FD), as well as by a Royal Society Research
Professorship (RP$\backslash$R1$\backslash$180112) (FD).

\bibliographystyle{JHEP} \bibliography{ngl}
\end{document}